\documentclass[a4paper,fleqn,usenatbib]{mnras}

\usepackage[T1]{fontenc}
\usepackage{ae,aecompl}

\usepackage{graphicx}	
\usepackage{amsmath}	
\usepackage{amssymb}	

\title[Pop III]{The Persistence of Population III Star Formation}

\author[R. H. Mebane et al.]{
Richard H. Mebane,$^{1}$\thanks{E-mail: rmebane@astro.ucla.edu}
Jordan Mirocha,$^{1}$
Steven R. Furlanetto$^{1}$
\\
$^{1}$Department of Physics \& Astronomy, University of California, Los Angeles, Los Angeles, CA 90095, USA
}

\date{Accepted XXX. Received YYY; in original form ZZZ}

\pubyear{2017}

\begin{document}
\label{firstpage}
\pagerange{\pageref{firstpage}--\pageref{lastpage}}
\maketitle

\begin{abstract}
We present a semi-analytic model of star formation in the early universe, beginning with the first metal-free stars. By employing a completely feedback-limited star formation prescription, stars form at maximum efficiency until the self-consistently calculated feedback processes 
halt formation. We account for a number of feedback processes including a meta-galactic Lyman-Werner background, supernovae, photoionization, and chemical feedback. Halos are evolved combining mass accretion rates found through abundance matching with our feedback-limited star formation prescription, allowing for a variety of Population~III (Pop~III) initial mass functions (IMFs). We find that, for a number of models, 
massive Pop~III star formation can continue on until at least $z \sim 20$ and potentially past $z \sim 6$ at rates of around $10^{-4}$ to $10^{-5}$ M$_\odot$ yr$^{-1}$ Mpc$^{-3}$, assuming these stars form in isolation. At this point Lyman-Werner feedback pushes the minimum halo mass for star formation above the atomic cooling threshold, cutting off the formation of massive Pop~III stars. We find that, in most models, Pop~II and Pop~III star formation co-exist over cosmological time-scales, with the total star formation rate density and resulting radiation background strongly dominated by the former before Pop~III star formation finally ends.
These halos form at most $\sim 10^3$ M$_\odot$ of massive Pop~III stars during this phase and typically have absolute magnitudes in the range of 
$M_\text{AB} = -5 $ to $ -10$. We also briefly discuss how future observations from telescopes such as JWST or WFIRST and 21-cm experiments may be able to constrain unknown parameters in our model such as the IMF, star formation prescription, or the physics of massive Pop~III stars.
\end{abstract}



\section{Introduction}

The first stars to have formed in the Universe were likely very different from those observed today, and their formation 
was crucial to the early evolution of galaxies. It is thought that these Population~III (Pop~III) stars formed in metal-free gas clouds through molecular hydrogen cooling, and they were therefore much more massive \citep[][]{bromm_1999, abel_2002} and luminous than today's metal-enriched stars (see \citealt{bromm_2013}). Because of their high luminosity and the small size of their birth halos, feedback may have played an important role in the formation of Pop~III stars \citep[][]{machacek_2001, wise_abel_2007, oshea_norman_2008, shapiro_2004}, and they most likely formed in very small numbers, perhaps in isolation \citep[][]{visbal_2017}. Despite their small numbers, they must have been able to produce enough metals to eventually allow their halos to begin forming the more traditional Population~II (Pop~II) stars in a reasonably short amount of time.

There have been many attempts to study the formation and properties of these Pop~III stars through the use of detailed numerical simulations (e.g., \citealt{machacek_2001}; \citealt{wise_abel_2007}; \citealt{oshea_norman_2008}; \citealt{xu_2016}; \citealt{stacy_2012}; \citealt{hirano_2015}), analytic arguments (e.g., \citealt{mckee_2008}), and semi-analytic models (e.g., \citealt{trenti_2009}; \citealt{jaacks_2017}; \citealt{visbal_2017}). Simulations of their supernovae (\citealt{heger_2002} and \citealt{heger_2010}) have found that such stars are able to produce a very high mass of metals that will eventually be used to cool gas more efficiently and form stars from a more traditional initial mass function (IMF). In some simulations, Pop~III stars are found 
to form often in binaries (\citealt{turk_2009}), which could potentially  
produce a cosmologically relevant abundance of Pop~III high-mass X-ray binaries.

While there has been much work done to study the properties of Pop~III stars, we have yet to directly observe a Pop~III halo. 
There has been one candidate at $z \sim 7$ described in \citet{sobral_2015}, although there are still many other possible explanations (e.g., \citealt{bowler_2017} and \citealt{agarwal_2016}), and recent ALMA observations have detected [CII] consistent with a normal star-forming galaxy \citep[][]{matthee_2017}. Unfortunately, since it is thought that Pop~III stars form in very low mass halos at high redshift, it may prove incredibly difficult to directly observe a Pop~III source. Even if they are too faint to observe with the James Webb Space Telescope (JWST), however, we may be able to detect their supernovae with an instrument such as the Wide-Field Infrared Survey Telescope (WFIRST) (\citealt{whalen_2013}), or we could potentially see their effect on the cosmological 21-cm background (\citealt{mirocha_2017}). All of these measurements
are very sensitive to the overall shape of the Pop~III star formation rate density and the timing of the transition to Pop~II star formation.

In order to understand how Pop~III halos make the transition to Pop~II star formation, we must focus on both internal and external processes. In a particular halo, star formation produces important feedback effects such as supernovae and photoionization \citep[see][]{bromm_2003, whalen_2008, whalen_2008b}. These effects can either limit or completely cut 
off Pop~III star formation based on the size of the halo and its growth due to mass accretion.

But the Pop~III phase is also sensitive to large-scale radiation fields generated by those stars and their successors. 
The evolution of Pop~III star formation in the presence of global feedback driven by Pop~II star formation is not well understood and is a key interest to this work. In particular, UV photons in the Lyman-Werner band emitted by Pop~II stars can destroy the molecular hydrogen in Pop~III clouds necessary for cooling (\citealt{haiman_1997}). 
This sets the minimum halo mass at which molecular hydrogen can self shield in a halo and cool to form stars (\citealt{visbal_2014}). This minimum mass is a crucial 
physical quantity, as the halo mass function is very steep. The properties of even the Pop~II halos at these redshifts are unknown, however, so studying their effects on the formation of Pop~III stars in a flexible model is of much use.

In this work, we employ 
an efficient and flexible semi-analytic model that allows us to model the formation of Pop~III stars over a wide range of parameters and assumptions. While our model is certainly simpler than numerical simulations with similar goals, we are able to compare many different models while still self-consistently computing the effects of a wide range of feedback processes including a meta-galactic Lyman-Werner background, supernovae, photoionization, and chemical feedback. As described below, we use many results from observations, simulations, and analytic arguments to simplify our model and justify our assumptions. 

Pop~III star formation is a complex process that can proceed through different channels. In this work, we will focus on the ``classical" process through which primordial gas remains neutral while it accretes onto a dark matter halo and then cools purely through H$_2$ line emission \citep[][]{bromm_1999, abel_2002}. The resulting star-forming regions are relatively hot and thus have Jeans masses of $\ga 100 \, M_\odot$, which likely leads to massive stars (though their actual masses are the subject of intense debate; e.g., \citealt{bromm_1999}; \citealt{abel_2002}; \citealt{clark_2011}; \citealt{hirano_2017}). This should be distinguished from metal-free stars that form after primordial gas is ionized, as occurs in massive halos with virial temperatures $\ga 10^4$~K: in that case, HD cooling is significant, which lowers the Jeans mass within the cooling clouds substantially 
\citep[][]{oh_2002, johnson_2006}. Because low-mass Pop~III stars do not differ dramatically from Pop~II stars \citep[][]{tumlinson_2000}, it is the first, massive generation that is of most interest. 
Here, we will therefore study how long this mode can persist through the early generations of galaxies.

In Section~\ref{sec:properties} we describe the properties of dark matter halos in our model,
including the halo mass function and the growth of halos.  In Section~\ref{sec:III} we discuss our treatment of the first halos which will form 
metal-free, massive Pop~III stars. In Section~\ref{sec:II} we describe the properties of the Pop~II halos which form the very first generations of metal-enriched stars.  We present our results in Section~\ref{sec:results}, caveats to our model in Section~\ref{sec:caveats}, a discussion of their implications for observations of star formation in the early universe in 
Section~\ref{sec:discussion}, and conclude in Section~\ref{sec:conclusion}.

In this work, we use a flat, $\Lambda$CDM cosmology with $\Omega_\text{m} = 0.28$, $\Omega_\text{b} = 0.046$, $\Omega_\Lambda = 0.72$, $\sigma_8 = 0.82$, $n_\text{s} = 0.95$, and $h=0.7$, consistent with the results from \citet{planck}. Any distances presented are shown in comoving units.

\section{Halo Properties}
\label{sec:properties}

Our model will follow the growth of a set of dark matter halos from very early times to $z=6$. To begin, we choose a set of halos with $z=6$ masses from $10^6 M_\odot$ to $10^{13} M_\odot$ in 10,000 logarithmically spaced bins. These halos are then tracked backwards to $z=50$ using the abundance matching technique described in section~\ref{sec:growth} in 1 Myr timesteps. With these mass histories in hand, we then track each halo individually, applying the semi-analytic model discussed in future sections to determine the feedback-limited star formation histories of both Pop~III and Pop~II stars. Global quantities (to be discussed in the following section) such as the star formation rate density and Lyman-Werner background are found by averaging all halos over the mass function.

\subsection{The Halo Mass Function}

The first component 
of our model is the number density of dark matter halos. We assume that all halos which can accrete gas will form stars, so knowing their abundance is of vital importance.

The halo mass function can be written as
\begin{equation}
n(M) = \frac{\bar{\rho}}{M} f(\sigma) 
\left| \frac{\text{d}\log{\sigma}}{\text{d}M}\right|,
\end{equation}
where $\sigma$ is the density variance, $\bar{\rho}$ is the matter density, and $f(\sigma)$ depends on the particular form of the mass function.  In general, $f(\sigma)$ can be found through fits to simulations or through analytic arguments. We test the dependence of our model on the chosen mass function by examining a Sheth-Tormen mass function \citep{sheth_tormen} and the mass function of \citet{trac_2015},  which is a fit to high redshift simulations. 
These two choices do not lead to any 
significant differences in our results, especially given the wide uncertainty in the other model parameters (see Fig.~\ref{fig:mass_histories}). We employ the use of the \citet{trac_2015} mass functions throughout the rest of this paper.

\subsection{The Growth of Dark Matter Halos}
\label{sec:growth}

There have been many attempts to model the overall mass accretion rates of dark matter halos following simulations or analytic arguments \citep[e.g.,][]{mcbride_2009, behroozi_2013, dekel_2013, goerdt_2015}.  For example, \citet{dekel_2013} write the overall mass accretion rate as
\begin{equation}
\dot{M} = A M^{\mu} \left( 1 + z \right)^{\beta},
\label{eq:mdot}
\end{equation}
where $M$ is the halo mass, $A$ is a normalization, $\mu \gtrsim 1$ from simulations (see \citealt{mcbride_2009} and \citealt{trac_2015}), and $\beta = 2.5$, which is consistent with analytic arguments \citep[e.g.,][]{neistein_2008}. 
By fitting to numerical simulations, \citet{trac_2015} found that $\mu = 1.06$ for $M \sim 10^8$--$10^{13} M_\odot$ at $z \sim 6$--10, which overlaps with the mass and redshift range we need but does not probe the full range we require.

While these accretion rates have been tested in simulations at moderate mass ranges and redshifts ($z \sim 6$), it is unclear that they are valid for the smallest halos which begin forming stars at very high redshift. Since the focus of our model is on Pop~III star formation, we 
use another model for the growth of halos.

\begin{figure}
	\includegraphics[width=\columnwidth]{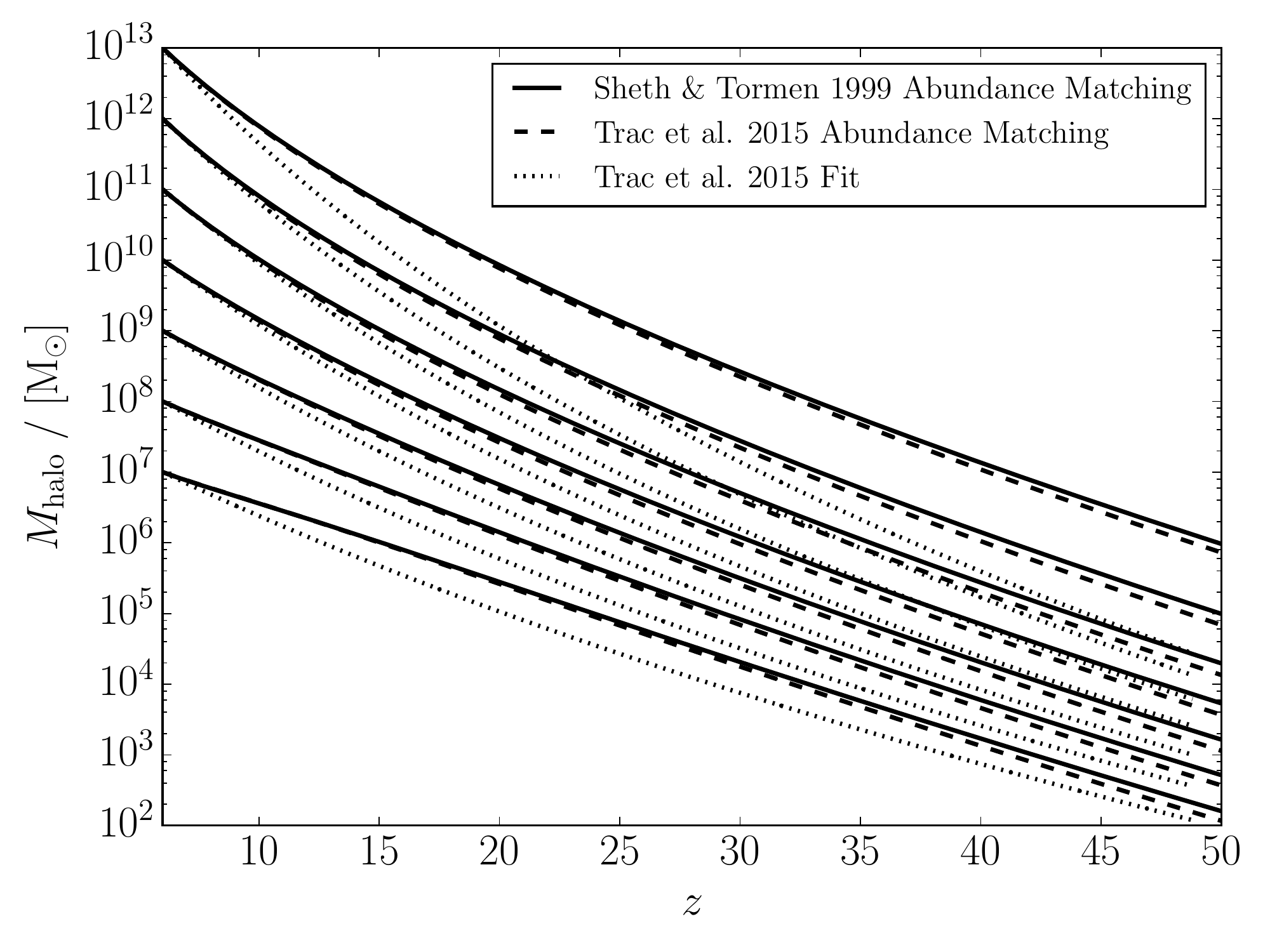}
    \caption{Mass histories of halos in our model. Halo growth is determined by abundance matching, where we assume halos remain at a constant comoving number density throughout time and find their mass by comparing mass functions at each time step (see section~\ref{sec:growth}). The solid and dashed lines 
    use mass functions from \citet{sheth_tormen} and \citet{trac_2015}, respectively.
    The dotted line is made using the \citet{trac_2015} fit to eq.~\ref{eq:mdot}. 
    Halos are initialized at the same mass at $z=6$, and their masses are then tracked backwards to $z=50$.     }
    \label{fig:mass_histories}
\end{figure}

We make the simple assumption that halos maintain their comoving number densities throughout time. This allows us to determine the mass histories of halos directly from the halo mass function by matching their abundances given by the mass function over a range of redshifts (see \citealt{furlanetto_2016} for a more detailed description of this technique). The mass histories for a number of halos in our model found through abundance matching are shown in Fig.~\ref{fig:mass_histories}.
Results with our two mass functions only differ by a factor of $\sim 2$ at the highest redshifts, so our analysis does not depend strongly on 
its exact form. 
The dotted curves use the \citet{trac_2015} simulation fit to eq.~\ref{eq:mdot}, which has a much larger discrepancy. However, 
it is only off by a factor of order unity for low mass halos and does not have a large effect on persistence of massive Pop~III star formation (see section~\ref{sec:trac_caveat}).

We note that our model for the growth of dark matter halos only follows the average growth, and thus we ignore the effect of mergers. \citet{behroozi_2015} found that the majority of accretion onto dark matter halos at high redshift occurs as smooth accretion from the IGM, though we will 
revisit the effects of mergers in section~\ref{sec:mergers}.

\section{Properties of Pop~III Halos}
\label{sec:III}

The first star-forming halos were very small and likely contained only a handful of very massive, metal-free stars.  These stars formed in molecular clouds that were cooled by molecular hydrogen, rather than the metal-line cooling which occurs in star-forming regions today.  Correctly modeling the growth of these halos and the time at which the first stars form is of vital importance to this work so that we can self-consistently model the feedback-limited star formation in the early universe and the transition to more typical, Pop~II star formation. In this section, we will describe the properties of these halos and the processes by which massive Pop~III star formation began.

\subsection{The Minimimum Mass}
\label{sec:min_mass}

In our model, we allow a new halo to begin forming stars only when it passes a minimum mass determined by the physics of H$_2$ cooling. 
\citet{tegmark_97} found that a halo must exceed a certain fraction of molecular hydrogen in order for cooling to become efficient, which is given by
\begin{multline}
f_{\text{crit, H}_2} \approx 1.6 \times 10^{-4} \left( \frac{1+z}{20}\right)^{-3/2}
\left( 1 + \frac{10 T_3^{7/2}}{60 + T_3^4}\right)^{-1} 
\\ \times \exp{\left(\frac{512 \text{K}}{T}\right)},
\label{eq:fh2}
\end{multline}
where $T$ is the virial temperature of the halo and $T_3 = T / 10^3$K.  In these halos, molecular hydrogren is formed primarily through the process
\begin{align}
\text{H} + \text{e}^- \rightarrow \text{H}^- + h\nu
\\
\text{H}^- + \text{H} \rightarrow \text{H}_2 + \text{e}^-,
\end{align}
where free electrons catalyze the reaction.  At high redshifts, H$^-$ can be easily destroyed by cosmic microwave background (CMB) photons, so we must balance the formation rate with the destruction rate to get the overall H$_2$ fraction in a halo.  \citet{tegmark_97} found this fraction to scale with a halo's virial temperature as
\begin{equation}
f_{\text{H}_2} \approx 3.5 \times 10^{-4} \text{ }T_3^{1.52}.
\end{equation}
Once $f_{\text{H}_2} > f_{\text{crit, H}_2}$, molecular hydrogen cooling will become efficient enough to cool gas into the first star-forming molecular clouds. In our model, we assume star formation begins immediately after this criterion is met, with no delay. This occurs in very 
small halos with masses $\sim 10^5$ $M_\odot$ at the highest redshifts in our model ($z \sim 50$).

After the first stars 
form, however, the minimum mass will instead be set by the metagalactic Lyman-Werner background. The Lyman-Werner (LW) band consists of photons in the energy range 11.5-13.6 eV that photodissociate molecular hydrogen through the Solomon process \citep[see][]{solomon}.  If a halo is present in a high enough background of Lyman-Werner photons, it can lose all of its molecular hydrogen and no longer be able to cool gas in star-forming regions.  Therefore, a self-consistent calculation of this background is required to determine the minimum masses of Pop~III halos 
\citep[see][]{haiman_1997, holzbauer_2012, visbal_2014}.

Following \citet{visbal_2014} the background can be written as
\begin{equation}
J_{\text{LW}}(z) = \frac{c}{4\pi} \int_{z}^{z_{\text{m}}}  
\frac{\text{d}t}{\text{d}z^\prime} \left(1+z\right)^3 \epsilon(z^\prime)
\text{d}z^\prime,
\label{eq:jlw}
\end{equation}
where $\epsilon(z)$ is the specific Lyman-Werner comoving luminosity density and $z_\text{m}$ is the maximum redshift at which Lyman-Werner photons can be emitted before they redshift into a Lyman line.  For simplicity, we assume that all photons can be redshifted by a maximum of 4\% \citep[][]{visbal_2014} before being absorbed by a Lyman resonance of neutral hydrogen in the IGM. This can be found from a simple fit to calculations of the redshifting of these lines.  We calculate $\epsilon(z)$ as
\begin{equation}
\epsilon(z) = \int_{M_\text{min}}^{\infty} n(M) 
\frac{\Omega_{\text{b}}}{\Omega_{\text{m}}} \frac{\dot{M}_\ast}{m_{\text{p}}} 
\left( \frac{N_{\text{LW}} E_{\text{LW}}}{\Delta \nu_{\text{LW}}} \right) 
 \text{d}M,
\end{equation}
where $M_\text{min}$ is the minimum mass at which a halo can form stars,
$\dot{M}_\ast$ is the star formation rate in a halo of mass $M$, $N_\text{LW}$ is the number of LW photons produced per baryon in stars, $E_\text{LW}$ is the average energy of a LW photon, and $\Delta \nu_{\text{LW}}$ is the frequency range of the LW band.  We take $N_\text{LW} = 1 \times 10^5$, which is 
roughly constant for massive Pop~III stars \citep{schaerer_2002}, $N_\text{LW} = 9690$ for Pop~II stars \citep{barkana_2005}, $E_\text{LW} = 11.9$  eV, and $\Delta \nu_{\text{LW}} = 5.8 \times 10^{14}$ Hz.  We compute $\dot{M}_\ast$ individually for each halo in both the Pop~II and Pop~III phases.

In order to determine which halos can form stars under a given LW background, we look to hydrodynamical simulations \citep[e.g.,][]{machacek_2001, wise_abel_2007, oshea_norman_2008}.  \citet{visbal_2014} find that the critical mass above which halos can form stars under a given LW background can be approximated by
\begin{equation}
M_{\text{min}} = 2.5 \times 10^5 \left( \frac{1+z}{26} \right)^{-1.5} 
\left( 1 + 6.96 \left( 4\pi J_{\text{LW}} \right)^{0.47}\right).
\label{eq:mcrit}
\end{equation}
Once we calculate $J_\text{LW}$ from our model, we can determine which halos will form massive Pop~III stars by only considering masses where $M > M_\text{min}$.

In this work, we neglect the potential feedback effects of X-rays, which can  catalyze 
H$_2$ formation by enhancing the free electron fraction and thus act as a source of positive feedback, or, could alternatively heat gas and prevent further fragmentation and star formation 
\citep[e.g.,][]{machacek_2003,kuhlen_2005}. The importance and sense of the feedback (i.e., positive or negative) will depend on the interplay between the 
Lyman-Werner and X-ray backgrounds, which in turn depends on the detailed properties of sources and their number density as a function of redshift \citep[e.g.,][]{ricotti_2016}. For now, we neglect these effects and defer a more detailed consideration to future work.

\subsection{The Pop~III Initial Mass Function}

Once a halo has exceeded the minimum mass required for Pop~III star formation, we begin to add the first stars. Since the gas clouds forming these stars cool primarily through molecular hydrogen, which is less efficient than metals, these kinds of Pop~III stars were likely very massive \citep{bromm_1999}. The Pop~III IMF is very uncertain, however, so we leave this as a parameter in our model (see Table~\ref{tab:popIII}). Stars are randomly sampled from the chosen IMF and placed in isolation in each star forming region.

The simplest IMF we consider is 
a delta function, where all Pop~III stars form at a single mass. The prefered mass is not known a priori, but one well-motivated choice is to follow \citet{mckee_2008}, who find the maximum mass of a massive Pop~III star by calculating the mass at which 
radiation pressure will stop accretion. 
\citet{mckee_2008} find this maximum mass to be
\begin{equation}
M_{\text{max}} \approx 145 M_\odot \left( \frac{25}{T_3}\right)^{0.24}.
\label{eq:mmax}
\end{equation}
If massive Pop~III stars form in isolation, then it is possible that all stars eventually reach this mass. In this case, we simply set the mass of every Pop~III star to this maximum mass. This case is especially interesting as 
it is just above the minimum mass at which pair-instability supernovae occur, so small details in the stellar masses are actually quite 
important.

We also allow for both 
lower mass and higher mass cases compared to the \citet{mckee_2008} model. In these cases we use a Salpeter-like IMF in various mass ranges. For the low-mass case, we use a minimum mass of 20 $M_\odot$ and a maximum mass of the \citet{mckee_2008} mass (see \citealt{bromm_1999} for a discussion of the fragmentation of protostellar disks and its effect on stellar masses). As an extreme high-mass case, we use a minimum mass of 200 $M_\odot$ and a maximum mass of 500 $M_\odot$. As we will discuss in section~\ref{sec:SN}, the choice of IMF has great implications for supernova feedback and its effects on these growing halos.

We note that our Pop~III model assumes a single star per site of star formation, with a mass drawn from the IMFs described above. This differs from other treatments \citep[e.g.,][]{jaacks_2017, visbal_2017}, which often utilize a fixed mass of star formation or a mass-dependent star formation efficiency. We compare to these similar works in section~\ref{sec:compare}, as well as test the dependence of our results on this assumption.

It is thought that these massive stars may often form in binaries \citep{turk_2009}, so we also investigate the case of a non-zero binary fraction. \citet{mirocha_2017} find that high-mass x-ray binary systems can produce a unique signature in the global 21-cm signal although, in this work, a nonzero binary fraction only serves to increase the overall star formation efficiency.

\begin{table*}
 \caption{Pop~III initial mass functions used in this work. Note that the metal yields are given as the mass of metals produced per unit mass of star formation.}
 \label{tab:popIII}
 \begin{tabular}{lllllll}
  \hline
  IMF & $M_\text{min}$ & $M_\text{max}$ & Slope & $E_\text{SN} / M_\ast$ [ergs / $M_\odot$] & C Yield & O Yield\\
  \hline
  Low Mass & 20 $M_\odot$ & $\sim 145 M_\odot$ (eq.~\ref{eq:mmax}) & $-2.35$ & $1.58 \times 10^{49}$ & $5.63 \times 10^{-3}$ &$6.25 \times 10^{-2}$\\[2pt] 
  Mid Mass & $\sim 145 M_\odot$ (eq.~\ref{eq:mmax}) & $\sim 145 M_\odot$ (eq.~\ref{eq:mmax}) & N/A & $6.90 \times 10^{49}$ & $3.13 \times 10^{-2}$ & $3.16 \times 10^{-1}$\\[2pt]
  High Mass & 200 $M_\odot$ & 500 $M_\odot$ & $-2.35$ & $1.41 \times 10^{49}$ & $1.41 \times 10^{-3}$ & $4.49 \times 10^{-2}$\\[2pt]
  \hline
 \end{tabular}
\end{table*}

\subsection{Photoionization Feedback from Pop~III Stars}
\label{sec:ion}

In order to determine how many Pop~III star forming regions should form in a single halo, we appeal to photoionization feedback. Since Pop~III stars were likely very massive and luminous, they would be able to ionize much of their surrounding material, making it unable to cool and form stars. Assuming a massive Pop~III star emits like a 
blackbody 
with effective temperature $T_{\rm eff}$ \citep[see][]{stacy_2012}, the number of ionizing photons produced is given by
\begin{equation}
\dot{N} = \frac{\pi L}{\sigma T_{\text{eff}}^4} 
\int_{\nu_{\text{min}}}^{\infty} \frac{B_{\nu}}{h\nu} \text{d}\nu,
\label{eq:ndot}
\end{equation}
where $L$ is the star's luminosity, $\sigma$ is the Stefan-Boltzmann constant, $\nu_\text{min}$ is the minimum photon frequency required for ionization, and $B_\nu$ is the blackbody spectrum of the star.  We approximate the luminosity of the star over its lifetime as the zero-age main sequence (ZAMS) luminosity, and we calculate $T_\text{eff}$ assuming the star emits like a blackbody.  We use the ZAMS stellar radius and luminosity given by \citet{stacy_2012} as
\begin{align}
L_{\text{ZAMS}}(M) = 1.4 \times 10^4 L_{\odot} \left(\frac{M}{10 M_{\odot}}\right)^2,
\\
R_{\text{ZAMS}} = 3.9 R_{\odot} \left(\frac{M}{10 M_{\odot}} \right)^{0.55}.
\end{align}

Once a massive Pop~III star is formed, it will begin to ionize the surrounding gas in its birth cloud.  If this radiation is able to ionize all of the gas available for star formation, then no more stars will form.  The size of the ionized region around a star is given by the Str\"{o}mgren radius, 
\begin{equation}
R_S = \left( \frac{3 \dot{N}}{4 \pi n^2_\text{H} \alpha_B}\right),
\end{equation}
where $\dot{N}$ is given by equation~(\ref{eq:ndot}) and $n_\text{H}$ is the hydrogen number density. \citet{bromm_2004} find that H$_2$ cooling tends to drive gas in these halos to densities of around 
$n_\text{H} \sim 10^4$ cm$^{-3}$ at $T \sim 200$ K. 

We then assume that the maximum number of stars allowed in a halo can be found by simply packing the halo with Str\"{o}mgren spheres. Under these assumptions, we find that a single star is able to ionize most of the gas in a halo at the masses and redshifts relevant to massive Pop~III star formation. If a halo is allowed to form Pop~III stars at higher masses ($\gtrsim 10^8$ M$_\odot$) it may be able to accomodate multiple star formation locations. Halos at these masses, however, are above the atomic cooling threshold and would therefore have already transitioned to Pop~II star formation in our model. We note that a single star is still able to ionize the majority of its surrounding gas even if we utilize a more complex model for the distribution of gas in a halo, such as the disc model described in \citet{munoz_2013}.

We note that \citet{whalen_2008} also find similar results for the destruction 
of H$_2$ in halos forming massive Pop~III stars. In their simulations, a 120 M$_\odot$ star is able to completely photodissociate all H$_2$ in a halo during its lifetime. As these stars form primarily through H$_2$ cooling, it is unlikely the gas in such a halo could cool and form stars near another massive Pop~III star.

\subsection{Pop~III Supernovae}
\label{sec:SN}

When a massive Pop~III star reaches the end of its life, taken to be 5 Myr in this model as stellar lifetimes do not vary 
appreciably at high mass \citep[][]{schaerer_2002}, the star will either explode in a supernova or collapse directly into a black hole. Stars with a mass between 40 $M_\odot$ and 140 $M_\odot$ and those with masses above 260 $M_\odot$ will not end their lives in a supernova, but they will rather 
collapse directly into a black hole.  Stars below this range will explode in a typical core-collapse supernova, while stars in the intermediate range  will end with a 
pair-instability supernova.  We therefore only take into consideration supernovae from stars which fall into these two categories. We assume that core-collapse supernovae release a kinetic energy of $10^{51}$ ergs, while pair-instability supernovae release $10^{52}$ ergs (see \citealt{wise_2008} and \citealt{greif_2010}).

Supernova feedback is particularly important in the early universe, as the typical kinetic energy released in a supernova could be orders of magnitude 
larger than the binding energy of the gas in a halo, which is of order 
\citep{first_galaxies}:
\begin{equation}
E_b \approx 2.53 \times 10^{50} \left(\frac{\Omega_\text{m}}{\Omega_\text{m}(z)} \right)^{1/3} 
 \left( \frac{M}{10^6 M_\odot} \right)^{5/3} \left( \frac{1+z}{10}\right) h^{2/3} \text{erg}.
 \label{eq:Eb}
\end{equation}

Every time a supernova explodes in a halo, we assume that the kinetic energy released is distributed throughout the remaining gas and calculate 
the corresponding temperature, assuming ionized, primordial gas. We then assume that the velocities of particles in this gas follow a Maxwell-Boltzmann distribution. 
This allows us to calculate the fraction of gas remaining in the halo and the fraction exceeding the halo's escape velocity. 
We then eject this gas, adding it to a reservoir that we later allow to re-accrete onto the halo over a time-scale equal to the halo's free fall time at the moment the gas is ejected,
\begin{equation}
t_{\text{ff}} = \frac{1}{4} \sqrt{\frac{3\pi}{2G\bar{\rho}_{\rm vir}}},
\end{equation}
where $\bar{\rho}_{\rm vir}$ is the average density of the halo (see \citealt{yoshida_2007} for a simulation of the re-incorporation of gas into a halo at these redshifts). 
However, if the velocity of the expelled gas is greater than $\sqrt{10} v_\text{esc}$, we assume that the gas has completely escaped and will never re-accrete. We discuss the importance of this assumption on our model in section~\ref{sec:reaccretion}.

This model for gas ejection has 
strong implications for a halo's transition to Pop~II star formation. We assume that the metals released by a supernova follow the gas, so the fraction of metals expelled is the same as the fraction of total gas expelled. This is discussed further in section~\ref{sec:retention}, where we investigate the fraction of metals which must be retained after a supernova in order for a halo to transition quickly.

\subsection{Transitioning to Metal-Enriched Star Formation}
\label{sec:trans}

One very important aspect of our model is the transition between metal-free, massive Pop~III star formation and metal-enriched Pop~II star formation.  The massive Pop~III stars we have discussed are likely born in only one star-forming region per halo (see section~\ref{sec:ion}), whereas Pop~II stars form from a more traditional, low-mass IMF (i.e., a Salpeter IMF).  In order to determine at what point halos switch to Pop~II star formation, we must look at how gas is cooled to form each type of star.  

Once a halo reaches a virial temperature of $T_\text{vir} = 10^{4}$ K, 
atomic line emission begins to dominate the cooling. Because of this, we assume that our treatment of massive Pop~III star formation is no longer valid,
\emph{even if the gas is still primordial}. Since 
atomic line emission is more efficient than H$_2$ cooling, gas clouds will be able to more easily fragment and form smaller stars. While these stars would still form from pristine gas, 
the stars' lower masses would cause the halo to be much more stable to supernovae feedback 
(so that the halo would transition to true Pop~II stars very soon anyway) and make their emission properties more like classical Pop~II stars than the very massive Pop~III phase.\footnote{We note that, for some of the more massive halos in our model (usually those greater than $10^{12} M_\odot$ at $z=6$), accretion rates can become higher than the metal production rate, which would cause a halo to never transition if not for this assumption. } 
We therefore treat this phase in the same way as we would a Pop~II halo.

Before a halo reaches the atomic cooling threshold, it can only transition to Pop~II star formation if it has retained enough of the metals released in supernovae. \citet{bromm_loeb_2003} find that the most important metal 
transitions to consider when looking at this 
critical point are CII and OI.  In particular, they find the 
metallicities at which metal line cooling will become more efficient than molecular hydrogen cooling to be [C/H]$_\text{crit} \approx -3.5$ and [O/H]$_\text{crit} \approx -3.05$, where [A/H] $= \log{(N_A / N_H)} - \log{(N_{A, \odot} / N_{H, \odot})}$.  Once a halo's mean metallicity surpasses one of these criteria, we then make the switch to the Pop~II star-forming regime.

We take our metal yields of Pop~III stars from \citet{heger_2010} and \citet{heger_2002}, which provide yields for core-collapse and pair-instability supernovae, respectively. Even though a single supernova may produce enough metals for the halo to reach the critical metallicity, this may not necessarily cause the halo to immediately begin forming Pop~II stars since not all metals will be retained. 
As discussed in section~\ref{sec:SN}, if metals are ejected from the halo, we must either wait until they re-accrete, until a later period of star formation where the halo is more stable to supernovae feedback, or until enough metals 
are retained in the presence of supernova feedback.\footnote{Note that we do not assume that the metals have thermalized; rather, we assume that they are entrained in the outflow from the halo.}
If the binding energy of the halo is small enough that the material is never re-accreted or the halo grows enough during the re-accretion time that the metal mass is insignificant, a halo may ``forget'' about its earliest periods of massive Pop~III star formation and require another burst of star formation before it can make the transition to metal-enriched stars.

Our 
model assumes that the mean metallicity of the halo is the deciding factor in the transition to Pop~II star formation, although this may not necessarily be the case. If metals can be more concentrated in certain regions of the halo, it is possible that Pop~II stars could begin to form even if the halo does not meet our critical metallicity requirements. We discuss the effects of this assumption in section~\ref{sec:retention} and leave a more detailed model of inhomogeneous mixing to a future work.

We also do not include the effects of external enrichment from nearby halos which could allow these massive halos to transition more quickly and also allow low mass halos to skip the Pop~III phase altogether (see \citealt{smith_2015}). \citet{jaacks_2017} find that the total metals produced in their model is not enough to raise the volume averaged metallicity in their simulations above the critical metallicity, indicating that, while a single halo may be able to produce enough metals to cut off star formation locally, it does not affect global Pop~III values.

\section{The Properties of Pop~II Halos}
\label{sec:II}

Since Pop~II stars form through the more efficient metal-line cooling rather than H$_2$, they will have much smaller masses that follow a more traditional IMF. In our model, we use a Salpeter IMF and break the assumption that stars form in isolation. We also stop tracking individual stars and simply 
prescribe the star formation efficiency in each halo. We assume a feedback-regulated star formation 
efficiency, as in \citet{furlanetto_2016} and \citet{sun_2015}, where star formation is regulated by supernovae through either energy or momentum conservation.

We note that our treatment of Pop~II star formation is much simpler than that of Pop~III stars. We are mostly concerned with the Pop~III stage of halo growth, so our Pop~II model is only used to calculate the radiation backgrounds which may contribute to limiting Pop~III star formation.

In our fiducial model, we find the star formation efficiency in a halo by balancing the kinetic energy released by supernovae with the binding energy of the halo. We follow \citet{furlanetto_2016}, who parameterize the star formation efficiency (defined here as the fraction of accreting material which turns into stars) as
\begin{equation}
f_\ast \approx \frac{1}{1 + \eta \left( M_h, z\right)},
\end{equation}
where $\eta$ relates the rate at which gas is ejected due to feedback to the star formation rate, $\dot{M}_{\text{ej}} = \eta \dot{M}_\ast$.

For energy-regulated 
star formation, we balance the rate at which 
supernova energy is released with the rate at which the binding energy of a halo is growing through newly accreted material,
\begin{equation}
\frac{1}{2}\dot{M}_\text{ej} v_\text{esc}^2 = \dot{M}_\ast \epsilon_\text{k} \omega_\text{SN}.
\end{equation}
Here, $v_\text{esc}$ is the escape velocity from the halo, $\dot{M}_\ast$ is the star formation rate, $\epsilon_\text{k}$ is the fraction of a supernova's kinetic energy used to drive a wind and lift gas out of the halo, and $\omega_\text{SN}$ is the kinetic energy released in supernovae per unit mass of stars. With this in mind, \citet{furlanetto_2016} find $\eta$ to be
\begin{equation}
\eta_{\text{E}} = 10 \epsilon_\text{k} \omega_{49} \left( \frac{10^{11.5} M_\odot}{M_h}\right)^{2/3} \left( \frac{9}{1 + z}\right).
\label{eq:etaE}
\end{equation}
Here, $\omega_\text{SN} = 10^{49} \omega_{49}$ erg $M_\odot^{-1}$ and $\epsilon_\text{k} = 0.1$ in our fiducial model.

We also consider the case where momentum is conserved, and we balance the star formation efficiency by comparing the momentum released in supernovae to the momentum required to eject gas from the halo. In this case, \citet{furlanetto_2016} find $\eta$ to be of the form
\begin{equation}
\eta_\text{p} = \epsilon_\text{p} \pi_\text{fid} \left( \frac{10^{11.5} M_\odot}{M_h}\right)^{1/3} \left( \frac{9}{1 + z}\right)^{1/2},
\end{equation}
where $\epsilon_\text{p}$ is the fraction of the momentum released in a supernova which drives the winds (taken here to be $\epsilon_\text{p} = 0.2$ in our fiducial model) and $\pi_\text{fid}$ defines the momentum injection rate from stars formed from a given IMF, defined as
\begin{equation}
\dot{P} = \pi_\text{fid} \dot{P}_0 \left( \frac{\dot{m}_\ast}{M_\odot / \text{yr}}\right).
\end{equation}
Here, $\dot{P}_0 = 2 \times 10^{33}$ g cm/s$^2$ and $\pi_\text{fid}$ is of order unity for a Salpeter IMF (see 
\citealt{furlanetto_2016} for a more detailed discussion of this model).

In both of these cases, we 
impose a maximum value on the fraction of accreted mass that will form stars, $f_\ast$. In our fiducial 
models, we assume $f_{\ast \text{max}} = 0.1$ in the energy-regulated case and $f_{\ast \text{max}} = 0.2$ in the momentum driven case as in \citet{furlanetto_2016}. We also assume that some fraction of newly accreted gas may be virial shock heated and be unavailable for star formation. We follow \citet{faucher_2011}, who find the fraction of accreted material that will be able to cool onto the halo and form stars to be
\begin{equation}
f_\text{cool} = 0.47 \left( \frac{1+z}{4} \right)^{0.38} \left( \frac{M_h}{10^{12} M_\odot} \right)^{-0.25}.
\end{equation}
This fraction is only less than 1 in the case of very massive halos at relatively 
low redshifts, so we find that this process does not have a large effect on our results.

\begin{figure*}
	\includegraphics[width=2\columnwidth]{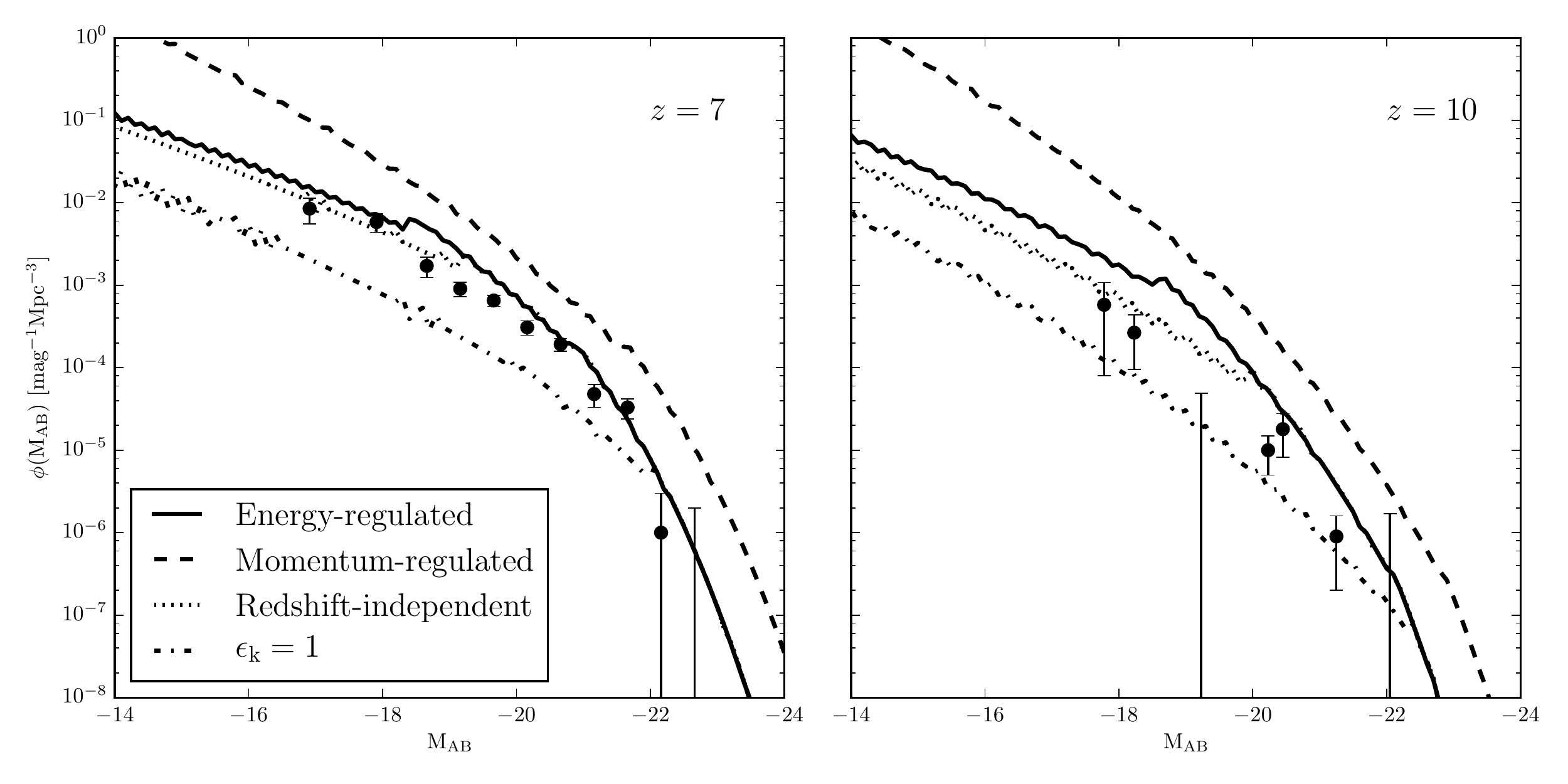}
    \caption{
    Luminosity functions for Pop~II stars in our fiducial model compared to the observed luminosity functions of \citet{bouwens_2015}. We do not include the luminosity functions for Pop~III stars as they are far too dim. In general, our results agree reasonably well with the data at $z=7$, but the energy- and momentum-regulated models overpredict the abundance of halos at $z=10$. We therefore also include a redshift-independent model 
    which fits the data better. Finally, we include an energy-regulated model with $\epsilon_\text{k}$ = 1 (i.e., all of the kinetic energy released by supernovae couples to a halo's gas to drive a wind).}
    \label{fig:LF}
\end{figure*}

We compare our results to to the observed luminosity functions of \citet{bouwens_2015} in Fig.~\ref{fig:LF}.\footnote{We note that other groups have produced luminosity functions at these redshifts \citep[e.g.,][]{mclure_2013, oesch_2013, schenker_2013, bowler_2015, finkelstein_2015} that are generally consistent with our results. The \citet{finkelstein_2015} data has a lower amplitude than our results and those of \citet{bouwens_2015}, although their shape is similar so this does not affect the model significantly as the amplitude is degenerate with our assumptions on the star formation efficiency. See \citet{mirocha_2016} and \citet{mason_2015} for more detailed comparisons.} We find reasonable agreement for our models at $z=7$, although our model overpredicts the abundance of halos at $z=10$. This was seen as well in \citet{furlanetto_2016}, who alleviated this discrepancy by including a redshift independent model of Pop~II star formation which fits the data better. We also include this model, which is identical to the energy-regulated model at $z=8$ with $\epsilon_\text{k} = 0.2$
but ignores the redshift dependence in 
equation~(\ref{eq:etaE}).

\section{Results}
\label{sec:results}

\begin{figure}
	\includegraphics[width=\columnwidth]{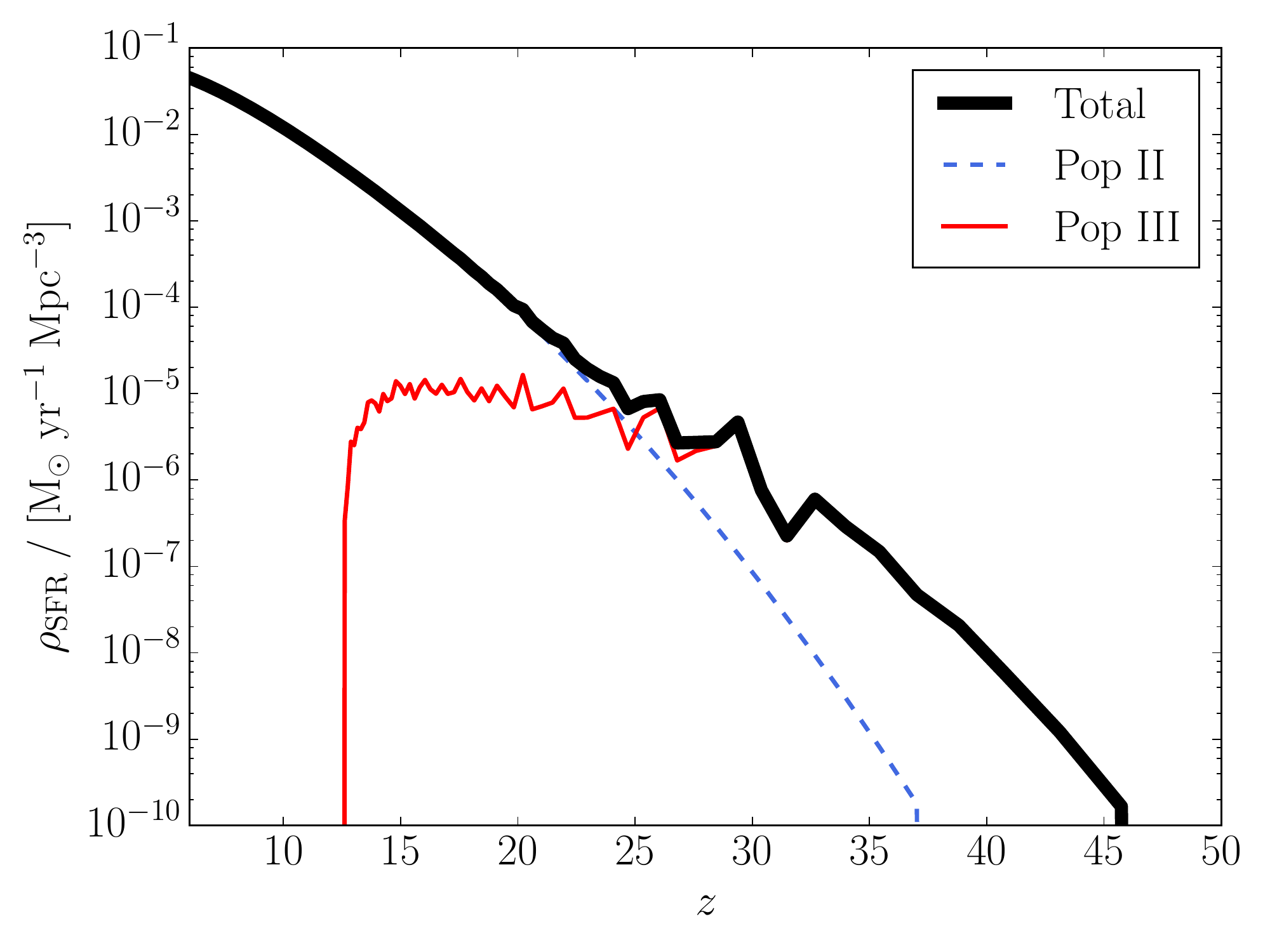}
    \caption{Star formation rate density of 
    massive Pop~III and Pop~II star formation in our fiducial model. Massive Pop~III star formation ends in our model once the minimum mass for star formation rises above the atomic cooling threshold (see Fig.~\ref{fig:min_masses_comp}).}
    \label{fig:sfrd_MT_fid_NLW}
\end{figure}

With all the components of our model in place, we now consider how our ensemble of halos evolves through the Pop~III era. 
In our fiducial model:
\begin{enumerate}
\item We assume a Salpeter IMF from 20~$M_\odot$ up to the limit in equation~(\ref{eq:mmax}) (henceforth referred to as ``low-mass" in comparison to more extreme models; see Table~\ref{tab:popIII}).
\item We set the gas re-accretion time equal to the halo free-fall time.
\item We assume that each star formation event generates just a single star or pair of stars, and we set the binary fraction to $0.5$.
\item We assume energy regulation (eq.~\ref{eq:etaE}) for the Pop~II phase.
\item We take the \citet{trac_2015} halo mass function and set the halo growth rate via abundance matching.
\end{enumerate}
We also consider a broad set of variations around these fiducial values. 

\begin{figure}
	\includegraphics[width=\columnwidth]{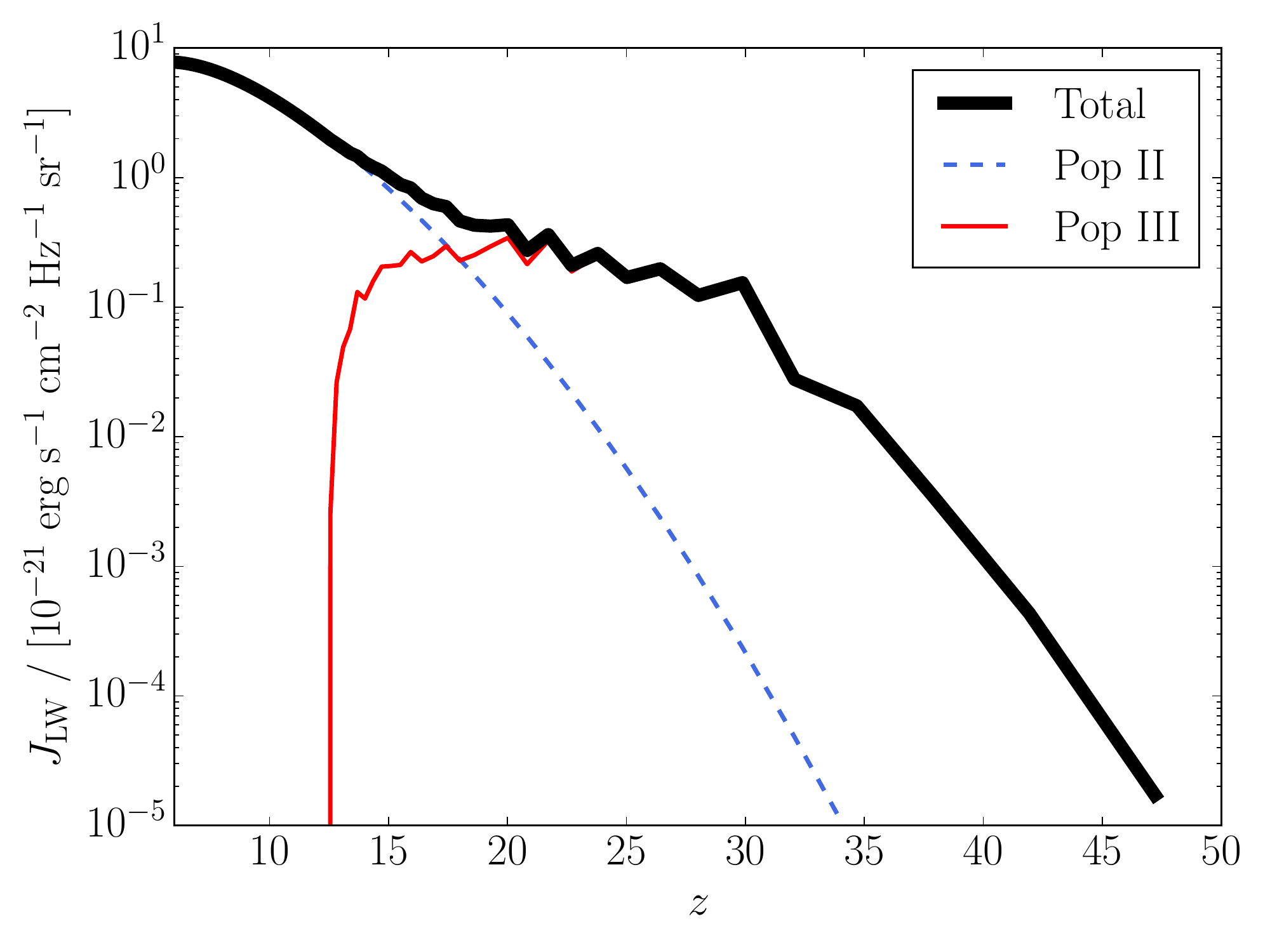}
    \caption{Lyman-Werner background for 
    our fiducial model. Note how the contribution from massive Pop~III stars is dominant for longer than the star formation rate densities in Fig.~\ref{fig:sfrd_MT_fid_NLW}. This is 
    because
    massive Pop~III stars are able to produce UV photons more efficiently than their Pop~II counterparts, so even a small amount of Pop~III star formation can continue to produce a high Lyman-Werner background.}
    \label{fig:jlws}
\end{figure}

Fig.~\ref{fig:sfrd_MT_fid_NLW} shows the total, 
massive Pop~III, and Pop~II star formation rate densities of our fiducial model. We see that 
massive Pop~III star formation 
gradually rises until $z \sim 25$, when the Lyman-Werner background has grown enough to narrow the allowed mass range of Pop~III halos sufficiently to begin the decline in star formation (see Fig.~\ref{fig:jlws}).  Note how 
massive Pop~III stars are able to dominate the Lyman-Werner background for a relatively long 
($z \sim 20$ instead of $z \sim 25$) time, even though Pop~II star formation has 
come to dominate the star formation rate densities much earlier (Fig.~\ref{fig:sfrd_MT_fid_NLW}). Since 
massive Pop~III stars
produce more UV photons per unit mass, small star formation rates can still create a large Lyman-Werner background. As this is happening, Pop~II star formation continues to rise as more Pop~III halos retain enough metals to make the transition, continuing until $z \sim 25$ where Pop~II star formation finally begins to dominate the total star formation in the Universe. After this, massive Pop~III star formation still continues on until $z \sim 12$, albeit at a much lower level than the total star formation. It is at this point that the minimum mass for massive Pop~III star formation rises above the atomic cooling threshold, cutting off the formation of any new Pop~III stars.

We next explore the origins of these features and their robustness in different scenarios of early star formation.

\subsection{Pop~III Star Formation in Individual Halos}

\begin{figure}
	\includegraphics[width=\columnwidth]{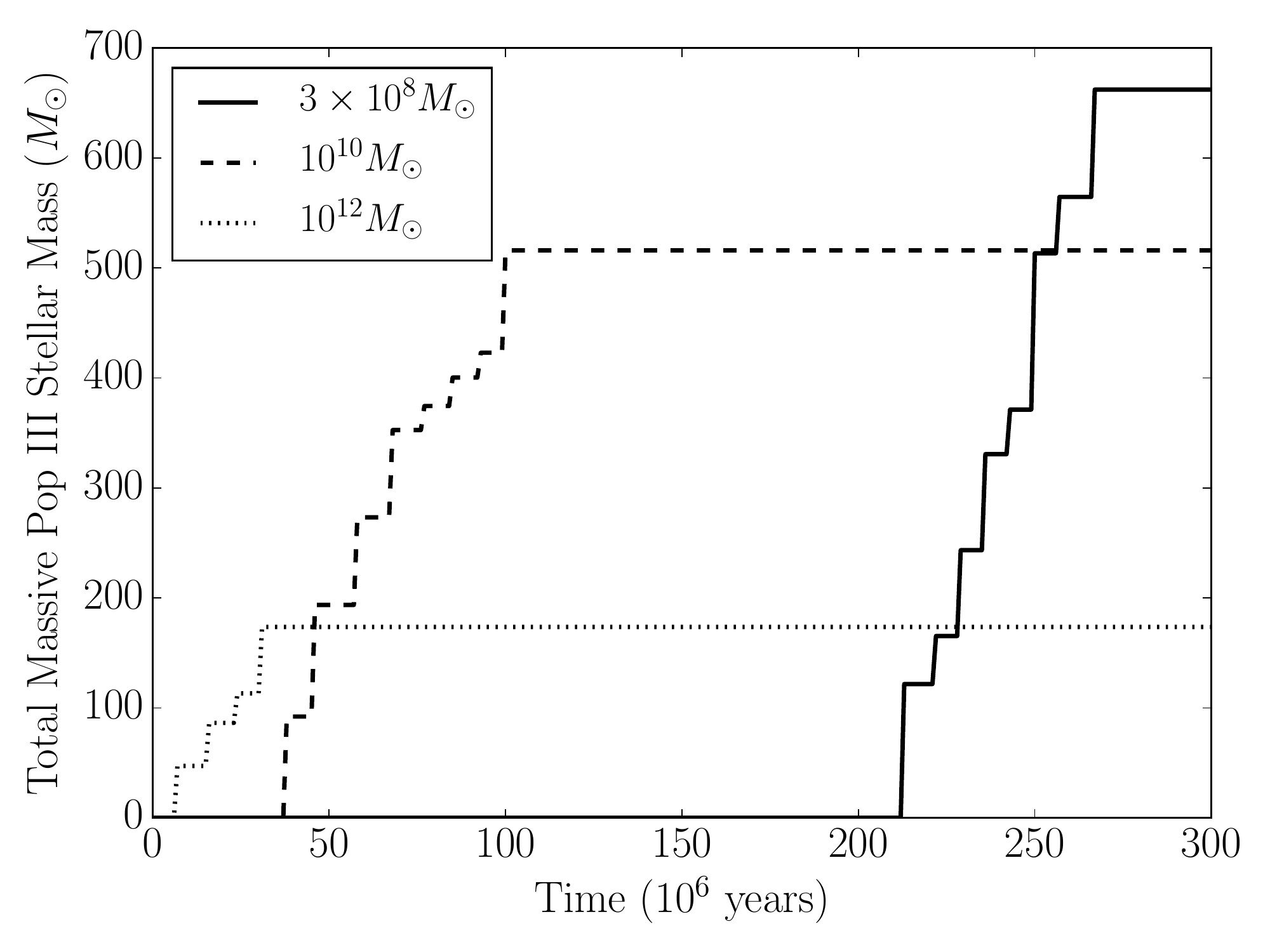}
    \caption{Total massive Pop~III stellar mass for three example halos of different masses. In general, less massive halos have more periods of star formation, which leads to a higher total mass produced. Masses shown are the final masses at $z=6$.}
    \label{fig:mass_tracks}
\end{figure}

To that end, we next consider the star formation histories of single halos. 
These are seen in Fig.~\ref{fig:mass_tracks}, which shows the total Pop~III stellar mass formed in 
example halos of three different masses. The burstiness of star formation is reflected in the discrete increases in the total mass when new stars are formed.  
Note that the halos which form latest -- and 
hence in our prescription have the smallest masses -- 
produce a \emph{larger} number of massive Pop~III stars than their 
more massive counterparts, even though the latter 
began forming their stars much earlier. This is due to the redshift dependence 
of the binding energy of a halo: 
at fixed halo mass, $E_b \propto (1+z)$ from eq.~(\ref{eq:Eb}).  
This allows 
a halo that forms earlier to become stable to supernovae feedback much 
more quickly, and it can therefore retain its metals after fewer periods of star formation.

\begin{figure}
	\includegraphics[width=\columnwidth]{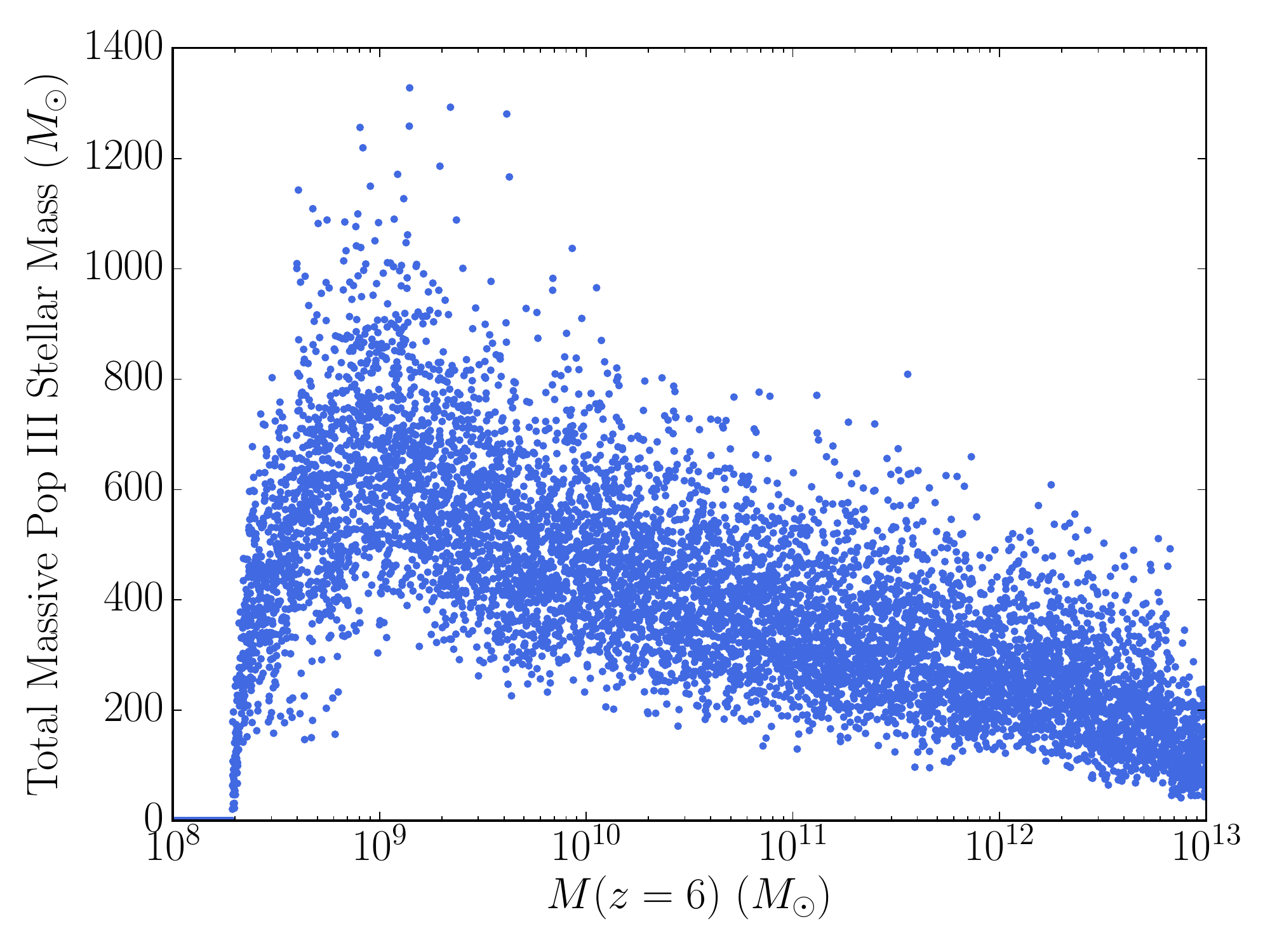}
    \caption{Total mass of 
     Pop~III stars formed in halos by $z=6$ as a function of mass. Each point corresponds to a halo in our fiducial model. Note that, since stars in our model form in isolation and will always die before the next period of star formation, this mass is not the total mass of Pop~III stars at $z=6$. Rather, this is the total mass that has formed in the halo, as most of the stars will 
     have either exploded in a supernova or collapsed to a black hole.}
    \label{fig:total_III_mass}
\end{figure}

Typically, halos go through no more than $\sim 10$ periods of 
massive Pop~III star formation, with a delay which depends on the accretion rate of the halo. In a halo with a smaller accretion rate, it may take a few million years to accumulate enough mass to form a single cloud above the local Jeans mass if supernovae from previous 
episodes have cleared out the halo. In halos with 
large accretion rates, however, star formation may be able to begin again with 
virtually no delay.
There is much stochasticity in this relationship, however, as can be seen in Fig.~\ref{fig:total_III_mass}. This is due to our Monte-Carlo treatment of the massive Pop~III star formation, where stars are randomly sampled from an IMF. If the IMF produces a significant number of stars which will not end their lives in a supernova (such as in the low-mass IMF model shown here), a halo can ``forget'' about earlier generations of Pop~III star formation as any metals produced will be lost to the black hole.

As noted in section~\ref{sec:trans}, these relationships would be very different if we did not allow halos to transition to Pop~II star formation at the atomic cooling threshold. In models where we do not 
impose this criterion, the most massive halos will spend the entire time ``stuck'' in the Pop~III phase
because their rapid accretion rates outpace the formation of metals. This causes the total mass of Pop~III stars produced to be much higher, as well as the star formation rate density of Pop~III stars to continue on until at least $z \sim 6$.

We find that halos in our model generally tend to become stable to supernovae shortly before reaching the atomic cooling threshold. In fact, most halos which transition by reaching the critical metallicity will end up crossing this threshold within the next $\sim$10 Myr, causing the minimum mass for Pop~II star formation to be very close to the atomic cooling threshold. Our results therefore generally support assumptions made in previous works that metal-enriched star formation will only occur in halos above the atomic cooling threshold.

Our model allows for a halo which has formed massive Pop~III stars in the past to fall below the minimum mass and cease star formation. This could, in principle, allow halos to stay dormant after a few earlier periods of star formation. In practice, however, we find that this does not happen, as halos tend to grow faster than the minimum mass, at least until the Lyman-Werner background has grown large enough to completely shut off massive Pop~III star formation.

\subsection{The Duration of 
Massive Pop~III Star Formation}
\label{sec:plateau}

The key question we wish to address in our models is how long massive Pop~III star formation persists under a variety of physics assumptions. Figure~\ref{fig:sfrd_comp} shows this for our suite of models. We find that, generically, the star formation rate density increases rapidly at early times before slowing dramatically or flattening. This ``plateau" period typically lasts multiple Hubble times before massive Pop~III star formation ends entirely. Extended Pop~III star formation such as this is also seen in 
the Renaissance Simulations \citep[see][]{xu_2016}.
 
The reason for the decline and end of massive Pop~III star formation is seen in Figs.~\ref{fig:jlws} and~\ref{fig:min_masses_comp}. As the Lyman-Werner background begins to build up at a faster rate 
when halos transition to Pop~II star formation, the minimum mass 
rises. As it gets closer to the atomic cooling threshold, the mass range in which halos are able to form massive Pop~III stars narrows, causing the star formation rate density to begin to plateau. Once the minimum mass rises above the atomic cooling threshold, massive Pop~III star formation ends. This is not a sign of ``self-regulation'' of Pop~III halos, however, as the Lyman-Werner background has become dominated by Pop~II stars at this point,
even though massive Pop~III stars are more efficient at producing Lyman-Werner photons. 

\subsection{The Importance of Pop~II Stars for 
Massive Pop~III Star Formation}

\begin{figure*}
	\includegraphics[width=2\columnwidth]{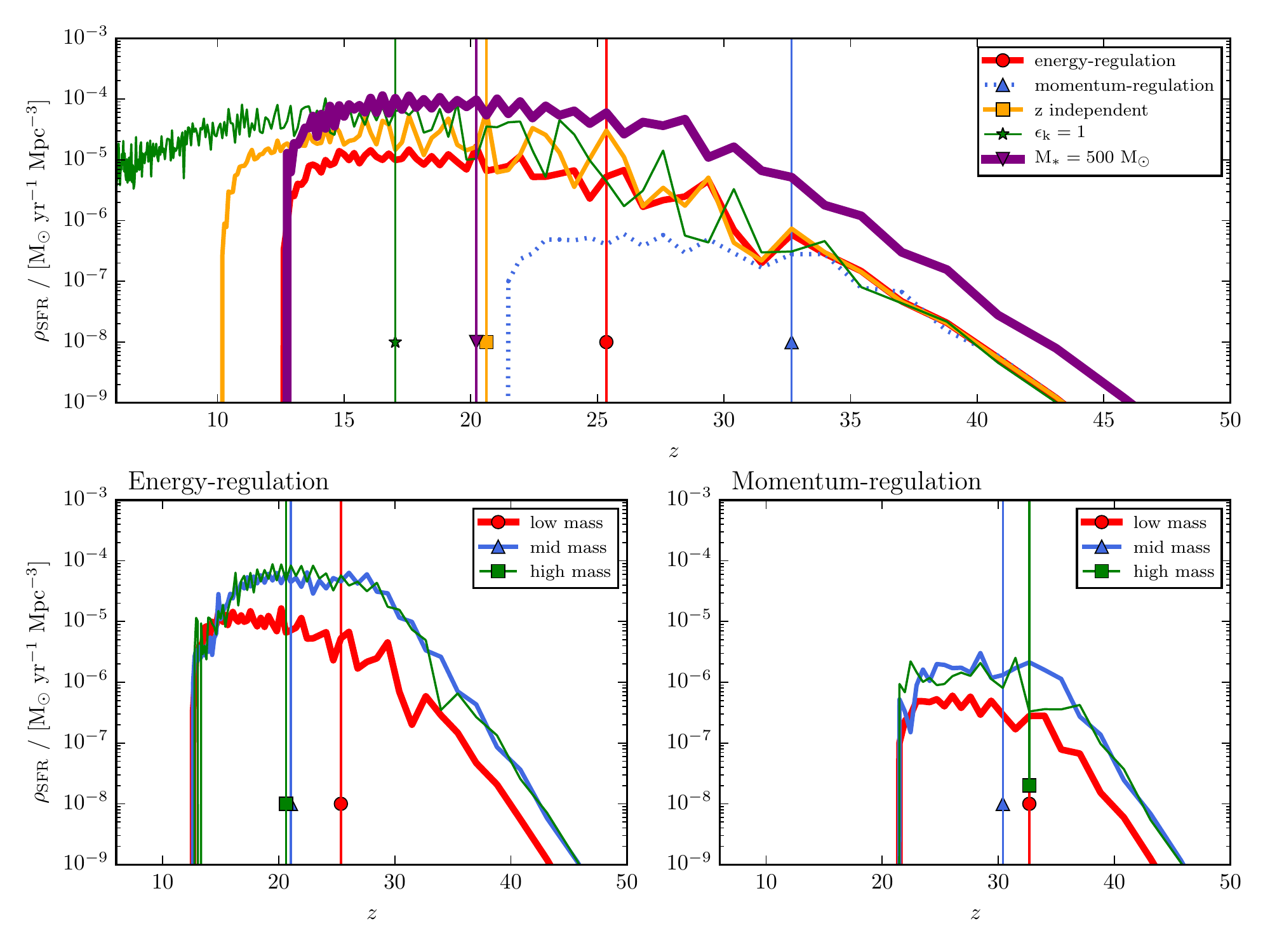}
    \caption{Star formation rate density of massive Pop~III stars for a variety of our models. Symbols indicate where Pop~II star formation overtakes Pop~III star formation. The upper panel shows our results for a low mass Pop~III IMF under a variety of different assumptions for the Pop~II and III star formation prescriptions. The bottom panels show a comparison between three different Pop~III IMFs using energy- and momentum-regulated Pop~II star formation, respectively. Note that models which employ momentum-regulated Pop~II star formation will form stars more efficiently in low-mass halos, raising the minimum mass above the atomic cooling threshold faster and cutting Pop~III star formation off sooner.}
    \label{fig:sfrd_comp}
\end{figure*}

\begin{figure*}
	\includegraphics[width=2\columnwidth]{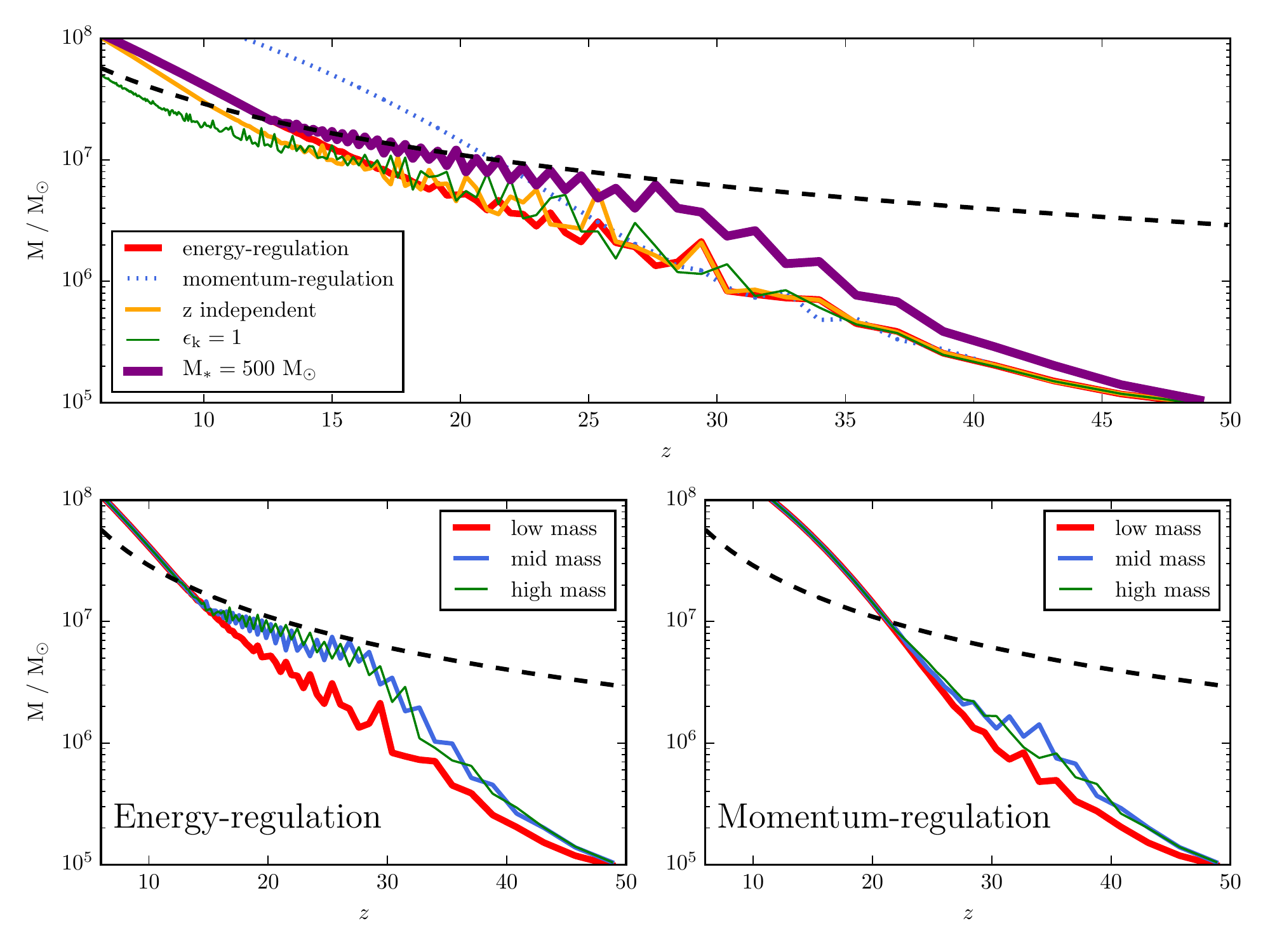}
    \caption{Minimum mass for Pop~III star formation for a variety of our models. The atomic cooling threshold is shown as the dashed line. The top panel shows our results for the low mass Pop~III IMF under a variety of different assumptions for the Pop~II and III star formation prescriptions. The bottom panels compare our results for different Pop~III IMFs using the energy- and momentum- regulated Pop~II star formation prescriptions. Once the minimum mass crosses the atomic cooling threshold, any new halos will begin forming low-mass stars, even if they form out of primordial gas. This is why the Pop~III star formation rate density vanishes so quickly in the momentum regulated models.}
    \label{fig:min_masses_comp}
\end{figure*}

\begin{figure}
	\includegraphics[width=\columnwidth]{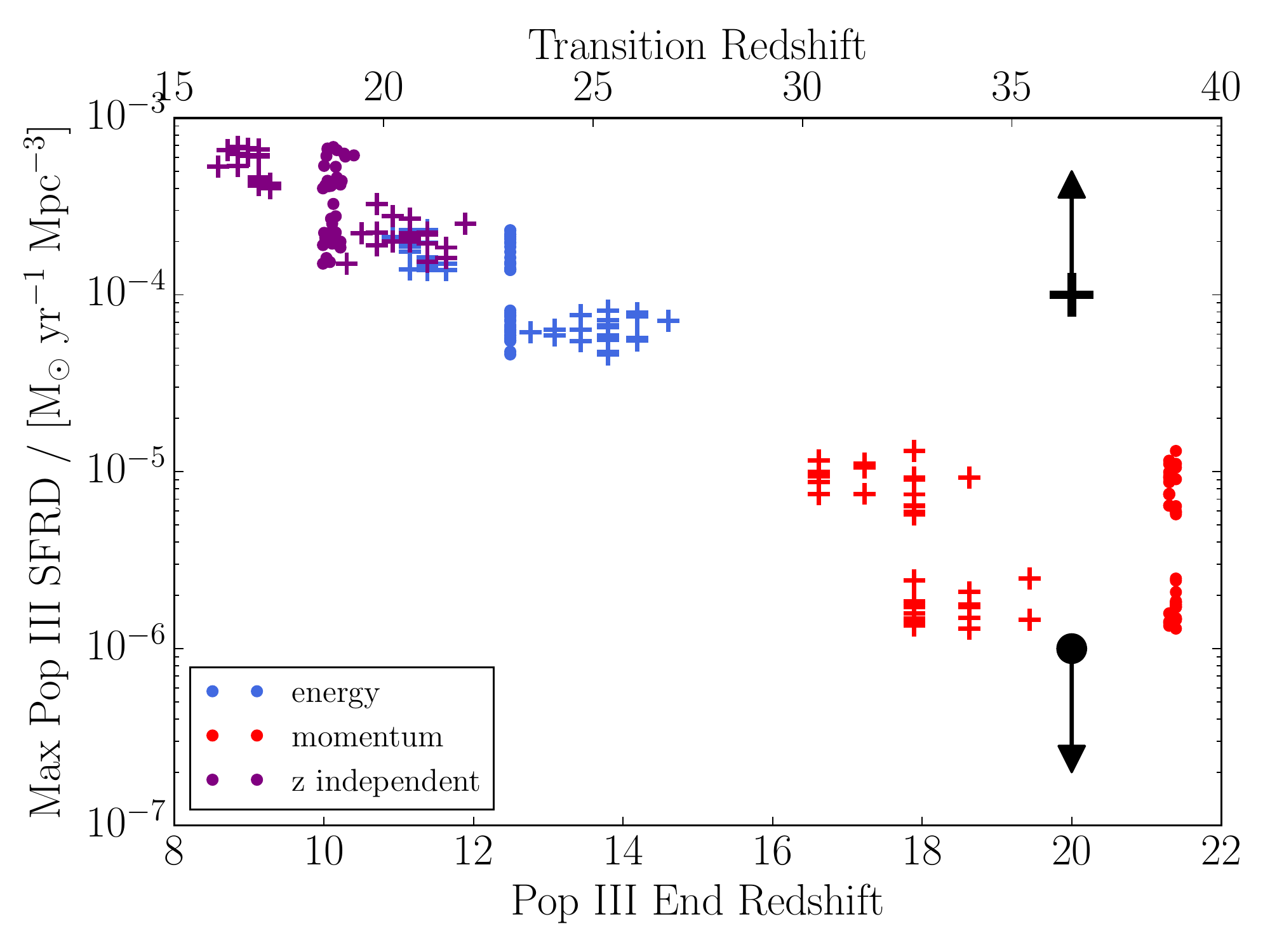}
    \caption{Maxmimum Pop~III star formation rate densities and important redshifts for a number of different Pop~II star formation prescriptions. Points correspond to the ending redshift of Pop~III star formation on the lower axis, and crosses correspond to the redshift at which Pop~II star formation overtakes Pop~III on the upper axis (indicated by the black symbols). Scatter for each model is caused by varying 
    IMFs as well as our Monte Carlo approach to model Pop~III star formation. Note how there is very little scatter in the Pop~III end redshift within each Pop~II model. By the time Pop~III star formation ends, the Lyman-Werner background is completely dominated by Pop~II star formation, so all models typically end at the same time.}
    \label{fig:scatter}
\end{figure}

As seen in the bottom panels of Fig.~\ref{fig:sfrd_comp}, the Pop~III star formation rate densities in models with momentum regulated Pop~II star formation are quite distinct from those with energy regulated Pop~II star formation. Because the star formation efficiency is higher in low mass halos (which are very abundant at these redshifts)  
under momentum regulation, the Lyman-Werner background and minimum mass of Pop~III halos rise much more quickly.  This is seen in the plots of star formation rate density as a much earlier transition to Pop~II star formation, which generally occurs at around the same time as the plateau. Also, since we force halos to transition to low-mass
star formation in our fiducial 
models once the minimum mass of Pop~III star formation crosses the atomic cooling threshold, all new halos will skip massive Pop~III star formation and begin forming Pop~II stars (see Fig.~\ref{fig:min_masses_comp}).

The IMF appears to have a more pronounced effect in the cases with energy-regulated star formation. In the case of the mid- and high-mass IMFs, the minimum mass follows the atomic cooling threshold before finally crossing it when Pop~II star formation is high enough. With more massive stars, the contribution to the Lyman-Werner background from massive Pop~III stars is larger, allowing these stars to more effectively regulate themselves near the atomic cooling threshold. Pop~III star formation does not entirely cease, however, until Pop~II stars contribute enough to the Lyman-Werner background to completely shut off star formation by themselves.

Figure~\ref{fig:scatter} illustrates these points for a wide range of model parameters. It shows two key transition points for several of our models. The filled circles, which should be read from the bottom axis, show $z_{\rm end}$, the times at which massive Pop~III star formation ends for each model. The crosses, which should be read from the top axis, show $z_{\rm II}$, the moments at which the Pop~II SFRD overtakes the Pop~III SFRD in each model. Within a given Pop~II model, there is very little spread in the redshift at which massive Pop~III star formation ends. This is due to the fact that, by the time this happens, the Lyman-Werner background is completely dominated by Pop~II stars. Fig.~\ref{fig:min_masses_comp} shows this as well: for a given Pop~II prescription, all the minimum mass curves cross the atomic cooling threshold at the same time, which marks the final endpoint of massive Pop~III star formation. However, there is quite a bit of scatter in the maximum star formation rate density at this time, because that depends more sensitively on our assumptions about the Pop~III stars. In fact, there appears to be two distinct ``clouds'' for each Pop~II star formation prescription in the plot. 
This is caused by our choice of IMF models, which separate (relatively) low- and (extremely) high-mass stars into contrasting cases. Because the prescriptions both form individual stars, the high-mass case produces about an order of magnitude more stellar mass per event, which directly affects the maximum star formation rate density. If we utilized a wider range of IMF models which uniformly spanned the relevant ranges in stellar mass, these ``clouds'' would be connected.

The transition redshift $z_{\rm II}$ also has a moderate amount of scatter, because it depends upon the amplitude of the Pop~III star formation rate density, which is sensitive to our assumptions. It is apparent that, in all of our models, the massive Pop~III era never has more than $\sim 10^{-3}$ M$_\odot$~Mpc$^{-3}$~yr$^{-1}$, and in most cases much less. 
This is comparable to the measured SFRD at $z \sim 10$ from \emph{bright} galaxies \citep[][]{zheng_2012, coe_2013, oesch_2014, mcleod_2015, atek_2015}, although in our models the total Pop~II SFRD is always much larger by $z \sim 10$. This illustrates the difficulty of detecting the extremely faint Pop~III halos, if they exist.

Figures~\ref{fig:sfrd_comp} and~\ref{fig:scatter} also show that the redshift independent case (which matches the observations best at $z=10$) yields a slightly more extended Pop~III star formation history, as it produces a smaller rate of star formation in low mass halos. As an extreme case, we also include in Figure~\ref{fig:sfrd_comp} a model with energy regulated Pop~II star formation, but with $\epsilon_\text{k} = 1$. In other words, all of the kinetic energy released by supernovae in this model is able to couple to the gas and work to lift it out of the halo.  Since feedback is stronger in this case, we see a smaller Pop~II star formation efficiency in halos, and therefore a lower Lyman-Werner background. As a result, the minimum mass to produce Pop~II stars never crosses the atomic cooling threshold. Because of this, massive Pop~III star formation is able to continue on until at least $z=6$. 
However, Figure~\ref{fig:LF} shows that this model substantially underpredicts the observed luminosity function at $z \sim 7$. We include it only to emphasize that the longevity of the massive Pop~III phase is very sensitive to the details of Pop~II star formation in low-mass halos.

\subsection{Self-Regulation of Pop~III Star Formation}
\label{sec:yields}

\begin{figure}
	\includegraphics[width=\columnwidth]{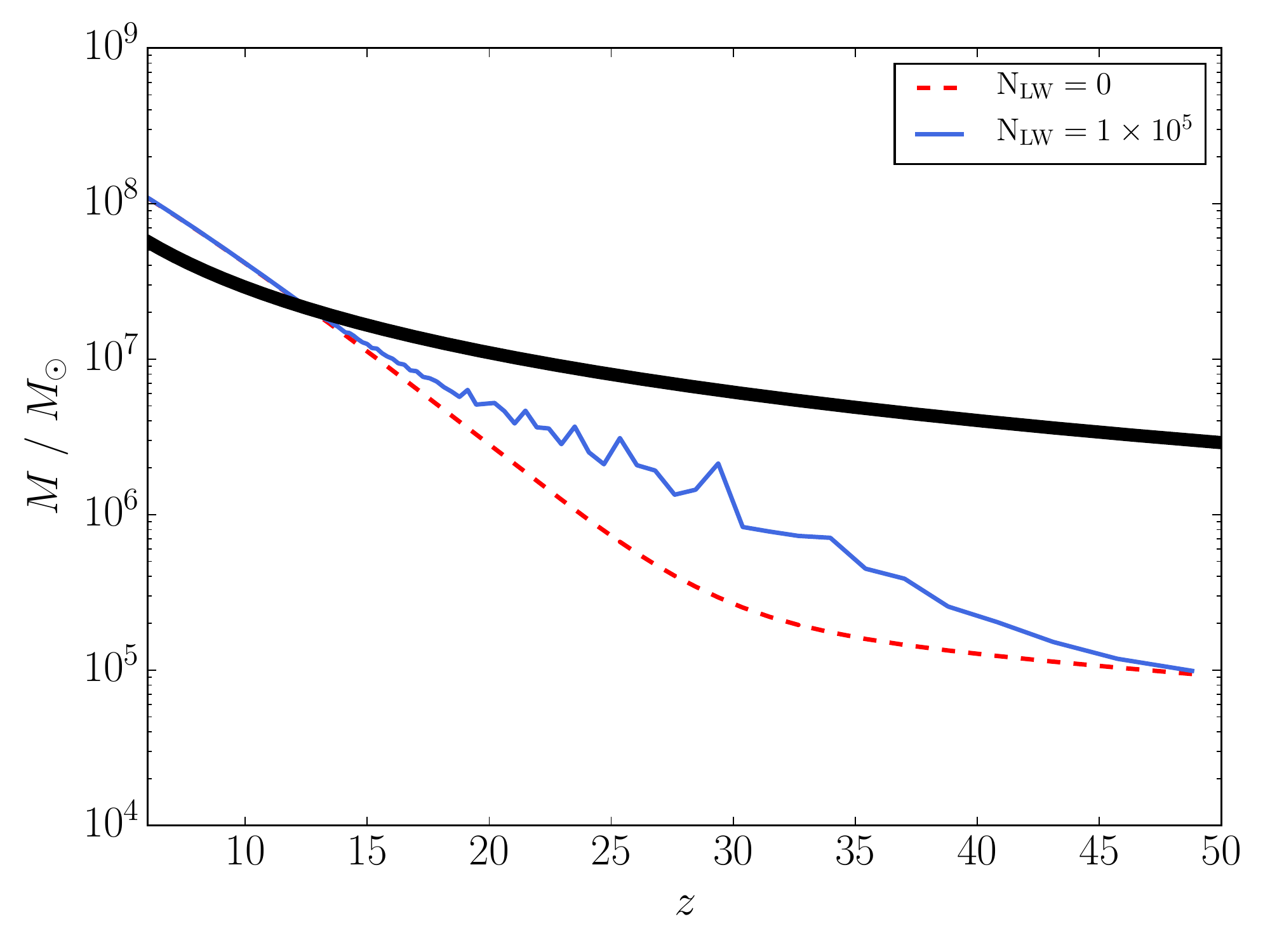}
    \caption{Minimum masses for Pop~III star formation for varying values of $N_\text{LW}$. The minimum mass is higher in the case where Pop~III stars contribute to the Lyman-Werner background, although all cases rise above the atomic cooling threshold at the same time. This indicates that global feedback from Pop~II star formation is really what ends the Pop~III phase in the universe.}
    \label{fig:min_mass_comp2}
\end{figure}

In order to test the 
ability of massive Pop~III stars to self-regulate themselves, we 
next consider a model in which Pop~III stars do not contribute to the Lyman-Werner background
by setting N$_\text{LW} = 0$ for Pop~III stars. The minimum mass of Pop~III halos for this case is shown in Fig.~\ref{fig:min_mass_comp2}. The Lyman-Werner background and therefore minimum mass will be lower at early times when Pop~III star formation is dominant. 
Without a Lyman-Werner background, halos form Pop~III stars earlier. But the time at which the minimum mass crosses the atomic cooling threshold is unchanged, because it is feedback from Pop~II stars which eventually causes 
Pop~III star formation to end.

\subsection{Comparison to Other Works}
\label{sec:compare}

Our model can be compared to similar semi-analytic models of Pop~III star formation which use slightly different approaches and study the effect of other parameters. 
For example, \citet{jaacks_2017} run a hydrodynamical simulation where Pop~III supernova remnants are ``painted'' onto the fluid 
with their properties calibrated from simulations and analytical arguments. They assume a fixed mass of $\sim 500$ M$_\odot$ per star formation event once a gas particle passes the thresholds of $n = 100$ cm$^{-3}$ with $T \leq 10^3$ K. 
They find, 
like us, that Pop~III star formation is not a self-terminating process. However, we come to this conclusion differently. They find that halos are simply not able to produce enough metals to raise the volume-averaged metallicity of their box above the critical metallicty required for Pop~II star formation. In our models, we find 
instead that massive Pop~III stars never form 
rapidly enough to raise the Lyman-Werner background high 
enough to completely cut off star formation 
in halos below the atomic cooling threshold on their own. We must instead wait until Pop~II star formation begins to dominate. We also find that, even if massive Pop~III stars do not contribute at all to the Lyman-Werner background, star formation will still end at the same time (see section~\ref{sec:yields}). They also find a maximum star formation rate density of around 
$10^{-3}$ M$_\odot$ yr$^{-1}$ Mpc$^{-3}$, however, which is about an order of magnitude higher than the star formation rate densities found in most of our models. 

In order to fully compare with their results, we include a model which also forms a fixed mass of $500$ M$_\odot$ per star formation episode. 
We note that, even under this assumption, we still find a star formation rate density about an order of magnitude smaller than 
theirs. 
This model is shown in Figs.~\ref{fig:sfrd_comp} and \ref{fig:min_masses_comp}. In this case, the minimum mass is higher before Pop~III star formation ends, because more stars form and the Lyman-Werner background is larger. However, the redshift at which massive Pop~III star formation ends is virtually unchanged, because by that point Pop~II star formation dominates by far. Thus this model can lead to a slightly larger Pop~III star formation rate density (though still about an order of magnitude smaller than that of \citealt{jaacks_2017}) but does not affect our major conclusions. Halos in this model tend to have a similar number of massive Pop~III star formation episodes as halos in our fiducial models, as the increased star formation results in more energy injected into the system from supernovae. While more metals are produced, they are much more easily ejected out of the halo, causing halos to be unable to transition due to reaching the critical metallicity. Thus, these halos tend to transition to Pop~II star formation simply by reaching the atomic cooling threshold.

The differences between our models and those of \citet{jaacks_2017} are mainly caused by different assumptions about the Pop~II star formation rate density. Their models have systematically lower star formation in the Pop~II phase than ours (their eq. 21), causing a lower Lyman-Werner background and therefore minimum mass. They also assumed lower values of $N_\text{LW}$ for both Pop~II and Pop~III stars (by factors of 5 and 10, respectively), exaggerating the decline in the Lyman-Werner background.

\citet{visbal_2017} apply a semi-analytic model to N-body dark matter simulations, allowing them to take into account any processes which require spatial information such as clustering and mergers. They allow stars to form at a specific fraction of a halo's baryonic mass, fiducially taken to be $10^{-3}$ (we find $\sim 5 \times 10^{-4}$ for the total massive Pop~III stellar mass formed in a halo in our fiducial, low-mass model). 
They find star formation rate densities consistent with ours, although their models only run to $z \sim 20$ so it is difficult to compare any results which rely on feedback from Pop~II stars, such as the duration of Pop~III star formation in the universe. They do find that the effects of external metal enrichment may be important only if metals are allowed to travel far from their original halos. This is similar to \citet{jaacks_2017}, who find that small halos which are externally enriched exhibit much lower metallicities than more massive halos which are internally enriched by their own star formation. While we do not include the effects of external enrichment in our model, we note that it would work to transition halos faster, potentially turning the plateau seen in many of our star formation rate densities into a more gradual decline.

\section{Caveats}
\label{sec:caveats}

In this section we describe some of the simplifications of our model and their consequences.

\subsection{Mass Growth Rates}
\label{sec:trac_caveat}

In Fig.~\ref{fig:mass_histories} we compare not only halo growth from our abundance matching technique, but also growth histories from the \citet{trac_2015} fit to eq.~\ref{eq:mdot}. In this case, halos which will end up with the same mass at $z=6$ will be less massive by a factor of order unity at $z=50$. These halos will therefore take slightly longer to cross the minimum mass and form their first massive Pop~III star. For example, in models using these accretion rates, the first Pop~III star will form at $z \sim 40$, compared to $z \sim 45$ in our fiducial models. Once Pop~III star formation begins to plateau, however, the two models become very similar. Because this assumption does not affect Pop~II star formation, the Lyman-Werner background is the same at later times, causing massive Pop~III star formation to end at the same time in both cases.

\subsection{Mergers}
\label{sec:mergers}

Our fiducial model assumes that halos grow primarily through smooth accretion from the IGM. While \citet{behroozi_2015} have shown that this is the primary source of growth for halos at high redshift, it is possible that mergers could also play an important role. In our model, the 
primary way in which mergers could change our results is if combining the metals produced in two merging halos allowed the halo to transition
to the Pop~II phase sooner than it would have on its own. However, we find that only a narrow range of halos are able to form massive Pop~III stars at any given time. 
Thus, with the exception of major events, only the larger mass progenitor would have been capable of creating stars in the past, 
so the smaller halo would have no metals to contribute. Even if it did, merging two halos currently forming Pop~III stars would still not cause a halo to exceed the critical metallicity unless one of the halos had already transitioned, as both the halos' gas and metals would be mixed. Because of this, we neglect the effect of mergers in our model. 
Nevertheless, we plan to investigate the effects of mergers in more detail in the future.

\subsection{Metal Retention}
\label{sec:retention}

\begin{figure}
	\includegraphics[width=\columnwidth]{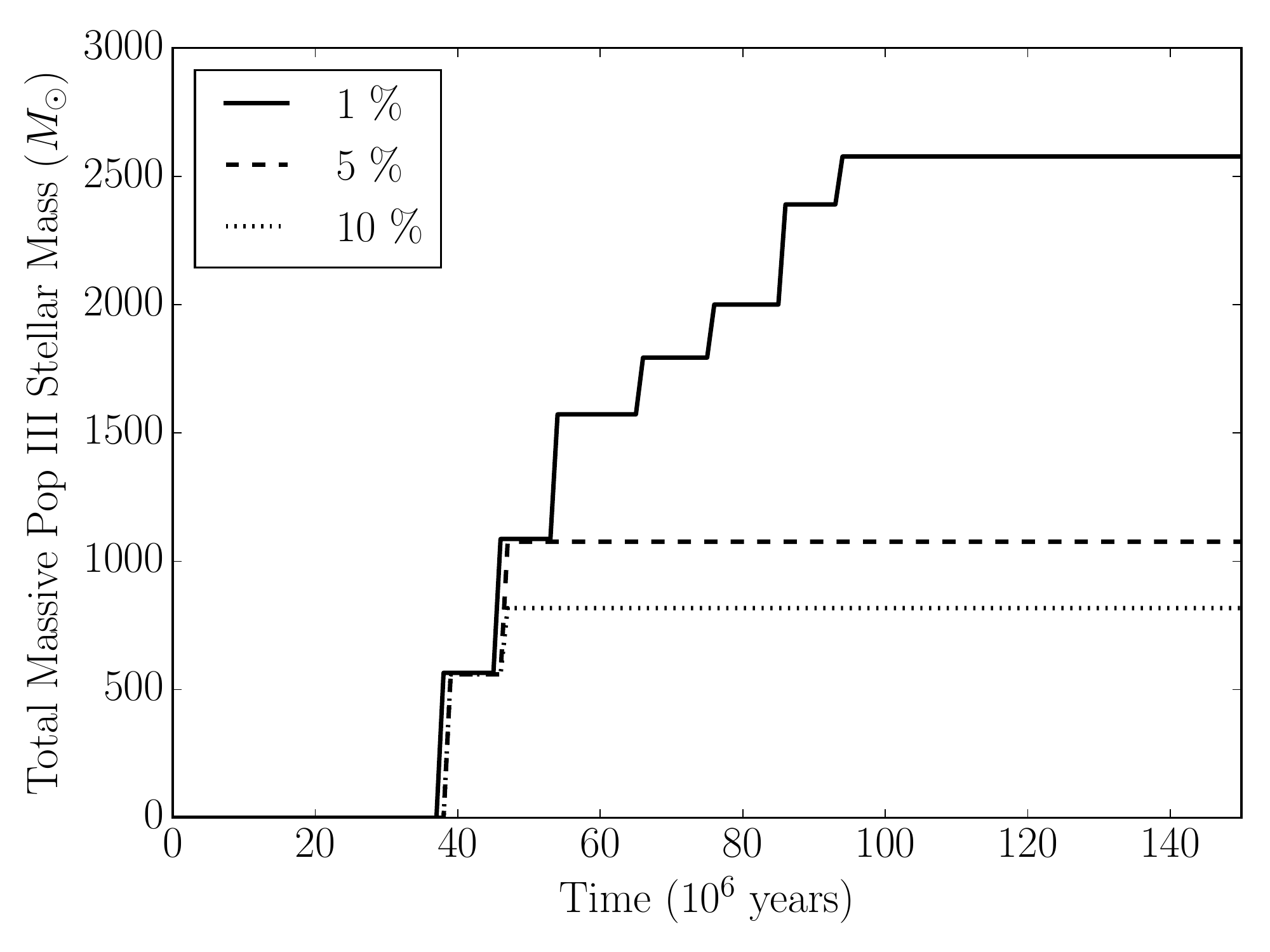}
    \caption{Star formation histories of a $10^{10} M_\odot$ halo with various metal retention fractions. Note that the results shown here are for the ``mid'' IMF in order to ensure every star will supernova and release metals.}
    \label{fig:retention}
\end{figure}

In order to test the importance of our Maxwell-Boltzmann treatment of the ejected gas as discussed in section~\ref{sec:SN}, we include an alternate prescription in which we fix the fraction of metals left behind after supernova feedback. Since Pop~III stars are able to produce such high masses of metals (see 
Table~\ref{tab:popIII}), we find that 
even if only a very small fraction ($\sim$5\%) of metals remain inside the halo after a supernova event, halos will \emph{immediately} transition to Pop~II star formation. This is shown in Fig.~\ref{fig:retention}, where a halo that retains 1\% of its metals is able to stay in the Pop~III phase for multiple episodes of star formation, while a halo with 5\% will transition much more quickly. This 
has the biggest effect in models which use a Pop~III IMF where every star ends its life in a supernova. In our low- and high-mass models, for example, some fraction of stars will directly collapse into a black hole, adding no metals to the halo itself. In this case, no matter how high the metal retention fraction is, a halo could go through many periods of massive Pop~III star formation if it happens to have formed a number of stars in mass ranges which do not produce supernovae.

\subsection{Gas Re-accretion}
\label{sec:reaccretion}

In our fiducial model, we assume that ejected gas re-accretes after one free-fall time, $t_\text{ff}$, calculated at the time the gas is ejected. In order to test this assumption, we ran models where this time was set to $1/10 t_\text{ff}$ and $10 t_\text{ff}$. In each case, we found results almost identical to our fiducial model. This indicates that re-accreted material is not important for the transition to Pop~II star formation, and halos which transition before the atomic cooling threshold do so when they become more stable to supernovae feedback and are able to retain metals produced in current periods of star formation.
This is because halos grow very quickly during the early phases of the Pop~III era, so they quickly transition from being so fragile that they are completely blown apart by a supernova (with no re-accretion) to being able to retain a fair fraction of their metals.

\subsection{Photoionization Heating}
\label{sec:photo}

Although we consider photoionization feedback inside each source's halo, assuming that it limits each halo to a single star forming region at any given time, we do not consider the effects of the photoionization on the gas surrounding the halo. Because the excess energy from ionizing photons typically heats gas to $\sim 10^4$~K, the resulting H~II regions will be much hotter than the average IGM. Even if the gas recombines, it will retain excess entropy, which will reduce the rate at which gas accretes onto the host halo \citep{oh_haiman_2003}. For simplicity, we ignore the potential of photoheating to suppress accretion onto the small halos in which our massive Pop~III stars form, which amounts to assuming that most of the accretion occurs through dense filaments that self-shield from the stellar radiation. If photoheating does suppress accretion, Pop~III halos will experience longer delays between star formation episodes. In the most extreme case, accretion would halt until the halos surpass a virial temperature of $\sim 10^4$~K, which we have shown is also approximately the point at which they become stable to supernova feedback. At that point, star formation will likely proceed similarly to our Pop~II prescription.

\section{Observational Implications}
\label{sec:discussion}

\subsection{Observing Pop~III Halos Directly}

\begin{figure}
	\includegraphics[width=\columnwidth]{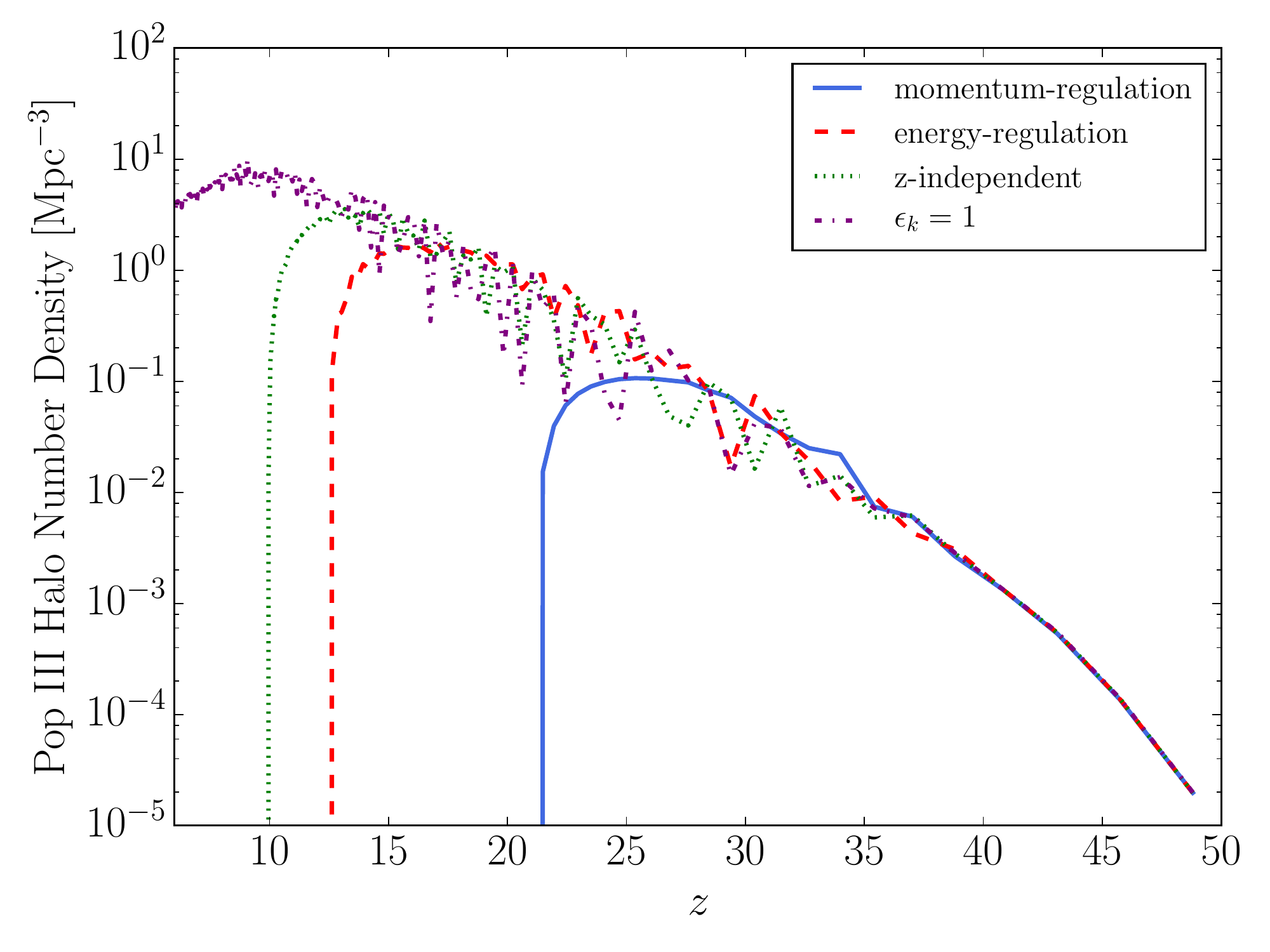}
    \caption{Number densities of Pop~III halos for a variety of models. This is calculated as the number of halos which are currently forming Pop~III stars (i.e, halos which have formed a Pop~III star in the past but have yet to make the transition to Pop~II star formation.)}
    \label{fig:pop_III_densities}
\end{figure}

Unfortunately, the luminosities of Pop~III halos in our model are very small and well below the capabilities of any current telescopes. 
We find that the absolute magnitude of these halos can vary between 
$M_\text{AB} \sim -5 $ for the lower mass Pop~III models to $\sim -10$ for the higher mass IMFs. Our models with a fixed mass of Pop~III stars are only slightly brighter, reaching $M_\text{AB} \sim -10.5$. While these halos are faint, though, they are actually quite abundant.
Fig.~\ref{fig:pop_III_densities} shows the number density of Pop~III halos for a variety of models. In the cases where Pop~III halos are around for the longest, their abundance is actually comparable to that of Pop~II 
galaxies (Fig.~\ref{fig:LF}).
Unfortunately, Pop~III halos in our model are far too dim to be detected by 
any forthcoming instruments, and they would likely require the use of lensing or an even more advanced generation of telescope to detect.

Observations of the luminous Ly$\alpha$ emitter CR7 by \citet{sobral_2015} have indicated the potential presence of a Pop~III halo at $z=6.6$ with a stellar mass of $\sim 10^7 M_\odot$. In order to find a halo with these properties in our model, we would have to break our single star-forming region assumption, as it is not possible for us to reach this mass with only a handful of massive Pop~III stars. We would also have to allow stars to form massive Pop~III stars in halos above the atomic cooling threshold, as it would otherwise be impossible for a halo to form such a high mass in stars (see Fig.~\ref{fig:min_masses_comp}). Recently, however, ALMA observations of this object have detected [CII] consistent with a normal star forming galaxy, so it is unlikely that CR7 actually contains $10^7$ M$_\odot$ in Pop~III stars \citep{matthee_2017}.

\subsection{
Massive Pop~III Supernova Rates}
\label{sec:SN_observe}

\begin{figure}
	\includegraphics[width=\columnwidth]{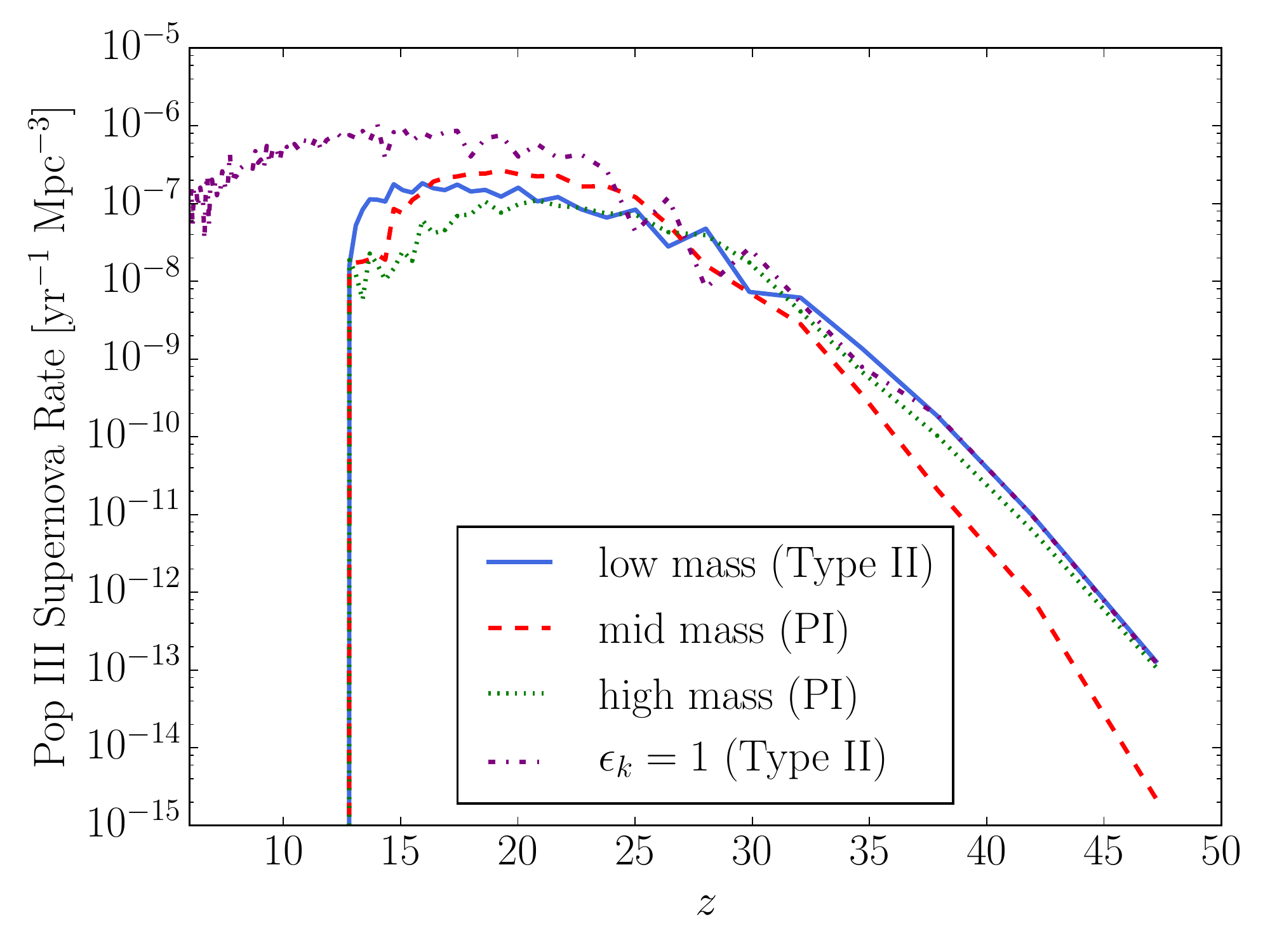}
    \caption{Pop~III supernovae rates for three of our fiducial models with energy-regulated Pop~II star formation. Note that the low mass model does not produce pair-instability supernovae. \citet{whalen_2013} find that these supernovae can be detected by JWST and WFIRST out to $z=30$ and $z=20$, respectively, which is where our calculated supernovae rates begin to flatten. The case with energy-regulated Pop~II star formation with $\epsilon_\text{k} = 1$ is shown as an example of a model which produces Pop~III supernovae out to at least $z=6$.}
    \label{fig:SN_rates} 
\end{figure}

While it is very unlikely that we will be able to directly observe a Pop~III halo in the near future, it may be possible to observe their supernovae. When a Pop~III star with a mass between $140 M_\odot$ and $260 M_\odot$ reaches the end of its life, it will likely explode in a pair-instability supernova. If Pop~III halos form many of their stars in this range (as in our model using 
\citealt{mckee_2008} masses or the high-mass, Salpeter-like IMF), then it may be possible to observe them with JWST or WFIRST. In particular, \citet{whalen_2013} find that Pop~III supernovae will be 
detectable out to $z=30$ for JWST and $z=20$ for WFIRST, which is in the redshift range in which our model 
produces the most supernovae. The Pop~III supernova rates from various models are shown in Fig.~\ref{fig:SN_rates}.
At $z=20$, an event rate of $\sim 10^{-6}$~Mpc$^{-3}$~yr$^{-1}$ translates to $\sim 3$ events per year per square degree per unit redshift. Thus, provided massive Pop~III stars produce luminous supernovae, these events may be
within reach of large-scale surveys.

\subsection{The 21cm Global Signal}

Pop~III stars will also affect the surrounding intergalactic medium (IGM) through their radiation fields.
The sky-averaged (``global'') 21-cm signal is a particularly appealing tracer of the IGM as it is sensitive not just to the ionization state, but the temperature and Lyman-$\alpha$ intensity as well. 
Its sensitivity to the thermal history of the IGM opens up the possibility that X-rays from 
Pop~III remnants could leave traces of their existence, in addition to the impact of 
Pop~III stars themselves.

In \citet{mirocha_2017},
we indeed find that the remnants of Pop~III stars have a unique impact on the signal. While in general the addition of new sources of X-rays reduces the contrast between otherwise cold neutral regions and the CMB and thus weakens the 21-cm background, Pop~III sources also give rise to a characteristic asymmetry due to the generic rise and fall of the Pop~III SFRD. In contrast, models neglecting Pop~III sources tend to be quite symmetrical \citep{mirocha_2016}.


\section{Conclusions}
\label{sec:conclusion}

We have presented a simple, semi-analytic model investigating the formation of
massive Pop~III stars in the early universe and the subsequent transition of their halos to the more traditional Pop~II star formation. Our model works by combining the results of a number of numerical simulations and analytic arguments with our self-consistent calculations of important feedback processes such as a meta-galactic Lyman-Werner background, supernovae, photoionization, and chemical feedback. From our results, we 
conclude that the star formation rate density of massive Pop~III stars increases rapidly as structure formation generates more halos at very high redshifts, until the stellar population increases enough to generate a substantial Lyman-Werner background, which slows the rate of star formation relative to halo formation. However, because Pop~III star formation is limited in each halo by chemical feedback, massive Pop~III stars are never able to self-regulate globally by generating a dominant Lyman-Werner background. Instead, more massive galaxies forming Pop~II stars are ultimately responsible for choking off massive Pop~III star formation in ``minihalos."

More specifically:
\begin{enumerate}
\item Depending on our choice of Pop~II star formation prescription, massive Pop~III stars can continue to form at a low level for an extended period of time, 
in principle until $z \sim 6$ at rates of around $10^{-4} - 10^{-5}$ M$_\odot$ yr$^{-1}$ Mpc$^{-3}$. In general, 
models with efficient star formation in low mass galaxies (i.e., our momentum-regulated model) will cut off massive Pop~III star formation much earlier by raising the minimum mass required Pop~III star formation to occur. Alternatively, inefficient star formation in low mass galaxies (i.e., our energy-regulated model) will allow massive Pop~III star formation to last longer.
\item The key parameters 
driving our results are the Pop~II star formation prescription and the Pop~III IMF. Secondary effects are the binary fraction, halo mass function, and Lyman-Werner yield of Pop~III stars.
\item Supernova
feedback is the most important feedback process in a single halo, 
because efficient expulsion of metals allows massive Pop~III star formation to persist in single halos for several generations. 
On a cosmological scale, the Lyman-Werner background 
dictates the halo masses at which Pop~III stars can form, and it is responsible for stopping the formation of new Pop~III stars once it causes the minimum mass to exceed the atomic cooling threshold.
\item While it may be possible to observe the presence of massive Pop~III stars through their supernovae or through cosmological 21-cm experiments, it is very unlikely that we will be able to directly observe a Pop~III halo in the near future. Our model produces Pop~III halos with magnitudes in the range of 
$M_\text{AB} = -5$ to $-10$ depending on our assumptions of the IMF, which is well below the capabilities of any current instruments.
\end{enumerate}

\section*{Acknowledgements}

This work was supported by the National Science Foundation through awards AST-1440343 and 1636646, by NASA through award NNX15AK80G, and was completed as part of the University of California Cosmic Dawn Initiative. In addition, this work was directly supported by the NASA Solar System Exploration Research Virtual Institute cooperative agreement number 80ARC017M0006. 
We also acknowledge a NASA contract supporting the ``WFIRST Extragalactic Potential Observations (EXPO) Science Investigation Team" (15-WFIRST15-0004), administered by GSFC. We acknowledge support from the University of California Office of the President Multicampus Research Programs and Initiatives through award MR-15-328388.

\emph{Software}: \textsc{matplotlib} \citep{matplotlib} and \textsc{numpy} \citep{numpy}




\bibliographystyle{mnras_edit}
\bibliography{refs} 








\bsp	
\label{lastpage}
\end{document}